\documentclass[journal = inoraj, manuscript=article,dvipsnames]{achemso} 
\setkeys{acs}{articletitle = true}
\usepackage{colortbl}
\usepackage{inputenc}
\usepackage{pdftexcmds}
\usepackage{titlesec}
\usepackage{amsfonts}
\usepackage{amssymb}
\usepackage{color,soul}
\usepackage{graphics}
\usepackage{graphicx}
\usepackage{gensymb}
\usepackage{cancel}
\usepackage{mathrsfs}
\usepackage{tikz}
\usepackage[version=4]{mhchem}
\usepackage{amsmath}
\usetikzlibrary{shapes}
\usepackage{booktabs}
\usepackage[table,xcdraw]{xcolor}
\usepackage{threeparttable}
\usepackage{pdfcomment}\usepackage{longtable}
\usepackage{booktabs}
 \usepackage{hyperref}
\usepackage{xr-hyper}   
\usepackage{lmodern}
\usepackage{textcomp}
\usepackage{multirow}
\usepackage[labelfont=bf]{caption}
\usepackage{subscript}
\usepackage{lscape}
\usepackage{gensymb}
\usepackage{bm}
\usepackage{titlesec}
\usepackage{etoolbox}
\usepackage[normalem]{ulem}
\usepackage{siunitx}
\usepackage[english]{babel}
\makeatletter
\patchcmd{\ttlh@hang}{\parindent\z@}{\parindent\z@\leavevmode}{}{}
\patchcmd{\ttlh@hang}{\noindent}{}{}{}
\makeatother
\usepackage{soul}

\newcounter{pevecounter} 
\providecommand{\PEVE}{abc}
\renewcommand{\PEVE}
  {\addtocounter{pevecounter}{1}\ifthenelse{\equal{\thepevecounter}{1}}{photo-electron valence-electron (PEVE)}{PEVE}}

\title{Revisiting the CeO Anion Photoelectron Spectrum: Resolving the Assignment and Photodetachment Mechanism of Satellite Transitions}

\author{Sakshi Nain}
\date{February 2026}
\affiliation[University of Louisville]
{Department of Chemistry, University of Louisville, 2320 S. Brook St., Louisville, Kentucky 40208, United States}
\author{Matheus M. F. de Moraes}
\affiliation[University of Louisville]{Department of Chemistry, University of Louisville, 2320 S. Brook St., Louisville, Kentucky 40208, United States}
\author{Hrant P.\ Hratchian}
\affiliation[University of California, Merced]
{Department of Chemistry, University of California, 5200 N. Lake Rd., Merced, California 95343, United States}
\author{Caroline C.\ Jarrold}
\affiliation[Indiana University Bloomington]
{Department of Chemistry, Indiana University, 800 E. Kirkwood Av., Bloomington, Indiana 47405, United States}
\author{Lee M.\ Thompson}
\affiliation[University of Louisville]{Department of Chemistry, University of Louisville, 2320 S. Brook St., Louisville, Kentucky 40208, United States}
\email{lee.thompson.1@louisville.edu}

\begin{document}
\begin{tocentry}
\includegraphics[height=4.5cm, width=4.5cm]{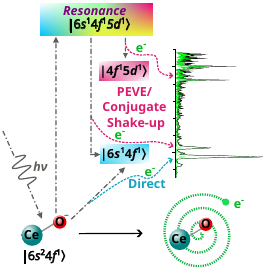}
\end{tocentry}

\newpage
\begin{abstract}
CeO serves as an important model system for understanding the interplay of electron correlation, spin-orbit coupling, and \(4f/5d\) configurational mixing in lanthanide oxides. Its anion photoelectron spectrum exhibits dense electronic manifolds and multielectron shake-up processes that complicate spectral assignments and photodetachment dynamics. Here, we combine high-resolution anion photoelectron spectroscopy with spin-free and spin--orbit-coupled CASSCF+NEVPT2 calculations to characterize the electronic structure of \ce{CeO^-} and CeO. A binning-based data analysis approach enables observation of weak transitions hidden by spectral congestion, while multireference calculations reproduce the low-lying electronic structure and detachment energetics. On this basis, tentative assignments are proposed for the observed photoelectron features. Analysis of photodetachment selection rules, photoelectron angular distributions, and multielectron detachment channels provides additional insight into the mechanisms underlying the spectrum. These results establish a comprehensive interpretation of the \ce{CeO^-} photoelectron spectrum 
and underscore the importance of combining high-resolution spectroscopy with multireference relativistic theory for electronically complex lanthanide systems.

\end{abstract}

\section{Introduction}
The unique electronic structure of lanthanide systems places them at the forefront of research, with applications spanning catalysis,\cite{shibasaki2002lanthanide} molecular magnetism,\cite{woodruff2013lanthanide,jin2026regulating} and light-emitting materials.\cite{li2025recent} Lanthanide oxides, in particular, provide an important platform for probing the electronic structure and magnetic properties of rare-earth elements.\cite{huizenga2021lanthanide} The exceptional magnetic anisotropy of many lanthanide ions, arising from localized $4f$ electrons and strong spin-orbit coupling (SOC), forms the basis of lanthanide-based single-molecule magnets (SMMs). The resulting ligand-field-split $4f$ manifold dictates both magnetic relaxation dynamics and magnetization-reversal barriers.\cite{marin2021shining,nain2024modulation} The strong electrostatic nature of oxide ligands creates well-defined ligand fields that dictate the ordering of low-lying electronic states.\cite{la2008theoretical} Lanthanide oxides thus provide a framework for understanding how metal–ligand interactions and SOC determine electronic structure and magnetic anisotropy.\cite{weichman2017electronic} In addition, lanthanide spectroscopy bridges fundamental electronic-structure theory and applications, as the 4$f$ manifolds generate distinctive, information-rich spectral signatures.\cite{coreno2006photoelectron,babin2021electronic} Within this context, simple lanthanide oxides serve as model systems to reveal how ligand-field effects, SOC, and orbital occupation determine low-lying electronic states. 

CeO, with its low 4$f$ occupation, represents an ideal platform for detailed characterization of electronic structure through absorption, emission, and laser spectroscopy.\cite{linton1983electronic} Its spectral features have also been detected in S-type stars, further stimulating interest in the study of cerium oxides.\cite{wyckoff1977ceo} The heavy cerium nucleus amplifies relativistic and spin-orbit effects, making CeO a valuable case study for investigating challenging multiconfigurational electronic structures.\cite{moriyama2013electronic} A reliable description of the low-lying manifold is therefore vital for spectral interpretation, bonding analysis, and method validation in $4f$- and $5d$-electron systems.

Anion photoelectron spectroscopy (PES) is a powerful tool for probing the low-lying electronic states of \ce{CeO^{-}}. However, several challenges complicate the complete resolution and assignment of spectral features.\cite{fischer2022photoelectron} One major challenge is the assignment of satellite peaks that are observed at higher binding energy and arise from shake-up processes, in which photodetachment of one electron is accompanied by simultaneous excitation of other electrons.\cite{simons2020ejecting,mosaferi2022interpretation} In lanthanides, the dense manifold of 4$f$ and 5$d$ configurations promotes electron correlation and configuration mixing, increasing the prevalence of shake-up excitations during photodetachment. Shake-up transitions probe multielectron interactions and 4$f$–valence-shell coupling. Their interpretation is complicated by unclear selection rules and intensities that often deviate from that typically observed in one-electron transitions.\cite{burroughs1976satellite,mason2019exceptionally} Consequently, understanding multielectron channels and their selection rules is critical for reliable spectral interpretation. It also provides quantitative insight into correlation effects, charge-transfer character, and the electronic structure that governs bonding and magnetic behavior in lanthanide oxides.\cite{teterin2004secondary} 

Theoretical modeling presents an additional challenge. The localized, multiconfigurational 4$f$ manifold and strong spin-orbit and relativistic effects often require treatments beyond conventional single-reference methods to accurately describe near-degenerate excited states. Multireference wavefunction-based methods are therefore required to capture both correlation and relativistic effects.\cite{zhou2022electronic,nain2024unravelling} Spectral analysis is further complicated by experimental constraints. The dense 4$f$/5$d$ manifold and finite signal-to-noise ratios often obscure weak transitions and hinder unambiguous spectral assignments.\cite{fischer2022photoelectron} Together, these challenges necessitate an integrated spectroscopic and computational approach. 

In this work, we build upon previous theoretical and experimental studies of CeO$^{-}$ and related lanthanide oxide systems to provide a more unified description of their low-lying electronic structure and photodetachment behavior.\cite{huizenga2021lanthanide,cao2021threshold} We present anion PE spectra with photoelectron counts binned in isoenergic intervals, which enables clearer observation of weak, one-electron forbidden transitions to higher-lying electronic states of the neutral. We further present a comprehensive computational study of \ce{CeO^-}, employing high-level multireference methods to accurately describe static and dynamic correlation within the \(4f\) and \(5d\) manifolds. Selection rules for shake-up states are developed and extended to provide a robust framework for interpreting multielectron detachment features in the photoelectron spectra. The photodetachment mechanism, including the observed photoelectron anisotropy parameters (\(\beta\)), is established through combined experiment and theory. Building on DFT-optimized geometries, spin-orbit-coupled potential energy curves (PECs) for the low-lying states of \ce{CeO^-} and CeO are computed at the CASSCF+NEVPT2 level. Analysis of the resulting transition energies, vibrational constants, and lifetimes in conjunction with the experimental spectra provides a detailed characterization of the electronic structure and photodetachment dynamics. This analysis highlights the interplay of 4$f$/5$d$ configurations, multiconfigurational mixing, and continuum effects in shaping the electronic structure of cerium oxide species.
\section{Methods}
\subsection{Photoelectron Spectroscopy Experimental Methods } \label{ssec:experimental}

The experimental PES measurements on CeO$^{-}$/CeO were presented in a previous study\cite{ray2015photoelectron} using an instrument described in detail elsewhere\cite{moravec1998study,waller2012study,felton2014measurement} but the experimental data have been processed in a different way than previously reported (see below). Briefly, CeO$^{-}$ anions were generated using a laser ablation cluster source, in which a solid Ce target was struck by the attenuated second harmonic output of a pulsed Nd:YAG laser (532 nm, 2.330 eV, 3 mJ/pulse), operated at a 30 Hz repetition rate, concurrently with pulses of ultra-high purity He, issued from a pulsed molecular beam valve.  The He pulse entrained the plasma generated by the ablation, and swept through a 2.5-cm long, 0.3-cm diameter clustering channel. The gas mixture expanded into a vacuum chamber and was collimated by a 3-mm skimmer. The anions were then accelerated in a time-of-flight mass spectrometer, where they separated in time and space per $m/z$.  Prior to colliding with a dual microchannel plate detector assembly, the anions were photodetached using a second Nd:YAG laser, timed to intersect only the 140 CeO$^{-}$ isotopomer ion packet. Both the second and third (355 nm, 3.495 eV) photon energies were used in this study. 

Anion PE spectroscopy is a mass-selective technique based on the photoelectric effect, in which a photon of known energy, h$\nu$, detaches the CeO$^{-}$ anion, resulting in neutral CeO in a distribution of rovibrionic states, and a photoelectron with kinetic energy (KE):
\begin{equation}
\mathrm{CeO^-}\left(E_{\mathrm{internal}}^{\mathrm{anion}}\right) + h\nu 
\rightarrow 
\mathrm{CeO}\left(E_{\mathrm{internal}}^{\mathrm{neutral}}\right) + e^{-}(\mathrm{KE})
\label{eq:photo_detachment}
\end{equation}
where $E_{\mathrm{internal}}^{\mathrm{anion}}$ and $E_{\mathrm{internal}}^{\mathrm{neutral}}$ represent the internal energies (electronic, vibrational, rotational) of the anion and neutral, respectively, though the technique generally lacks rotational resolution.  If the anion is internally cold, $E_{\mathrm{internal}}^{\mathrm{anion}} \approx 0$, the electron kinetic energy (eKE) distribution reflects the energies of the neutral rovibronic levels accessed via photodetachment:
\begin{equation}
\mathrm{eKE} = h\nu - \mathrm{EA} - E_{\mathrm{internal}}^{\mathrm{neutral}} + E_{\mathrm{internal}}^{\mathrm{anion}} 
\label{eq:KE}
\end{equation}
The eKE distribution is measured from the drift times of the photoelectrons that happen to travel the length of a 1-m, field-free drift tube situated perpendicular to the ion drift path (0.02\% collection efficiency), at the intersection of the ion beam and the laser interaction region. The drift times are converted into eKE, calibrated by setting common transitions observed in the spectra obtained using different photon energies to the same eBE. Spectra measured with different photon energies will exhibit common transitions appearing at different eKE values (eq.\ \ref{eq:KE}). We therefore present spectra in terms of electron counts versus electron binding energy, eBE, where 
\begin{equation}
\mathrm{eBE} = h\nu - \mathrm{eKE}
\label{eq:BE}
\end{equation}
is independent of photon energy, and is the energy of the final neutral state relative to the initial state of the anion.  The spectral linewidth of the instrument used in this study\cite{felton2014measurement}
\begin{equation}
\Delta \mathrm{eKE} = 0.004\,\mathrm{eV} + 0.0078\,\mathrm{eV} \left(\frac{\mathrm{eKE}}{\mathrm{eV}}\right)^{3/2}
\label{eq:}
\end{equation}
is narrower for features observed at lower eKE, so transitions measured with lower photon energy are better resolved than the same transition measured with higher photon energy. Because the laser interaction region and the electron drift tube are field-free, the photoelectron angular distributions (PAD) can be measured by rotating the laser polarization relative to the electron drift path. The PAD, which reflects the symmetry of the molecular orbital associated with the detachment transition, can be expressed by
\begin{equation}
I_{\theta} \propto a\left[1 + \beta P_{2}(\cos\theta)\right]
\end{equation}
where $I_\theta$ is the intensity of photoelectrons ejected at the angle $\theta$ relative to the electric field vector of the detachment laser, $\alpha$ is a simple scalar, $P_2(cos\theta$) is the second-order Legendre polynomial, and $\beta$ is an eKE -dependent anisotropy parameter that is related to the character of the molecular orbital. $\beta$ can be approximated directly from the relative intensities of electrons ejected parallel to ($I_\parallel$) and perpendicular to ($I_\perp$) the electric field vector of the laser via,
\begin{equation}
\beta = \frac{I_{\parallel} - I_{\perp}}{\tfrac{1}{2} I_{\parallel} + I_{\perp}} \label{eq:beta}
\end{equation}
Spectra were collected in 40,000 laser shot increments, alternating between parallel and perpendicular laser polarization to avoid changes in relative photoelectron yields due to fluctuations in the ion signal intensity.  Spectra are typically accumulated for 500,000 to 1,000,000 shots.

The electron drift times are recorded using a digitizing oscilloscope that measures scintillations produced from a constant fraction discriminator, which creates a consistent pulse for each electron, triggered by the amplified signal from a second microchannel detector assembly positioned at the end of the 1-m electron drift tube.  The oscilloscope is triggered by the output of the detachment laser (via a photodiode), and records the thus-standardized electron scintillations for 4 microseconds, in 1.6 ns bins. The standardized electron signal is $ca$. 5 ns wide.  In converting from drift time to eKE, each increment in eKE corresponds to a diverging window of drift time.  The data presented below have their electron counts binned into equal 0.002 eV increments.  Electron yields at the high eBE (low eKE) portion of the spectrum are therefore integrated over a wider window of time.  This process was not applied to the previously published data.  As a result, previously published spectroscopic features in the higher eBE range appear lower in intensity than the same features presented below, but the former has many more data points (associated with 1.6 ns time intervals) in the energy interval of the feature.

\subsection{Computational Details}
Initial molecular geometries for CeO and CeO$^{-}$ have been obtained from scalar-relativistic DFT optimizations and subsequently employed as starting points for computing the multireference PECs (Section S1 in supporting information). 
Complete active space self-consistent field (CASSCF) calculations have been performed to capture static correlation, followed by second-order N-electron valence state perturbation theory (NEVPT2) for dynamic correlation.\cite{malmqvist1989casscf,angeli2001introduction} The spin–orbit interactions are included using quasi-degenerate perturbation theory (QDPT) in combination with the spin–orbit mean field (SOMF) operator.\cite{hess1996mean,angeli2004quasidegenerate} For CeO, the active space has been systematically increased from a minimal CAS(2,8) including Ce 4$f$ and 6$s$ orbitals, to an inner valence CAS(8,16), adding Ce 5$d$, 6$p$ and O 2$p$ orbitals (Section S2 in SI). Given the minimal changes in low-lying state ordering across the tested active spaces, the final CAS$(n,13)$ space for CeO$^-$ ($n=3$) and CeO ($n=2$), comprising Ce \(4f\), \(5d\), and \(6s\) orbitals, is adopted as the optimal balance between accuracy and computational cost and used for all subsequent calculations. 
The scalar relativistic effects are included via the second-order Douglas–Kroll–Hess (DKH2) Hamiltonian, combined with DKH-compatible def2-TZVP basis sets for both Ce and O.\cite{weigend2005balanced,nakajima2012douglas} 

An energy window of approximately 22,000 cm$^{-1}$ has been employed to select the number of electronic states in the CASSCF, ensuring inclusion of all relevant low-lying states for subsequent SOC analysis. A total of 77 quartet, 95 doublet states for \ce{CeO^{-}}, and 42 triplet, 42 singlet states for neutral CeO are included in the state averaged CASSCF treatment, ensuring an adequate description of the low-lying electronic manifold. All the relativistic quantum chemical calculations have been performed with ORCA 6.1.1.\cite{neese2025software} The excitation energies of the low-lying electronic states 
are obtained from CASSCF+NEVPT2 calculations at both the spin-free and spin–orbit-coupled levels. 

\section{Results and Discussion}
The spectral intensities and angular distributions measured in anion photoelectron spectroscopy provide a direct probe of the underlying photodetachment dynamics. We first outline the governing principles of molecular photodetachment, including selection rules, angular distributions, and multi-electron effects influencing the anisotropy parameter ($\beta$). These concepts are then applied to the spectrum of \ce{CeO^{-}} to assign transitions and elucidate the electronic structure of \ce{CeO}.
\subsection{Photodetachment Theory}
Decades of research on anion photodetachment and molecular photoionization have established a robust theoretical framework for interpreting photoelectron spectra.\cite{reid2003photoelectron,sanov2014laboratory} Within this framework, spectral intensities and photoelectron angular distributions provide detailed insight into electronic structure and detachment dynamics.\cite{anstoter2021modeling,tufekci2024anion} In molecular photodetachment, these observables are governed by electronic-state symmetry, continuum-electron dynamics, and detachment selection rules. Their interpretation can be further complicated by multielectron excitations and relativistic effects, particularly spin-orbit coupling. The following sub-sections summarize the theoretical principles connecting these effects to the anisotropy parameter $\beta$, which form the basis for interpreting the \ce{CeO^-} photoelectron spectrum.

\subsubsection{Selection Rules}   \label{sec:selection_rules}
The selection rules for photodetachment from a diatomic molecular orbital were first expressed by Xie and Zare.\cite{Xie.1990} While these selection rules are rigorously based on considerations of changes in quantum numbers for the possible photodetachment channels, they do not provide insight into the detachment orbitals involved. Moravec and coworkers later discussed how quantum number selection rules place restrictions on the detachment orbital in \ce{Sn2-}.\cite{Moravec.1999} Here we express the analysis of Xie and Zare in a second-quantization formalism and extend it  through a Dyson orbital decomposition analysis to provide a mechanistic interpretation of the detachment process in terms of the allowed detachment orbitals.

The intensities of transitions in the photoionization spectrum are determined by consideration of the remnant molecule combined with the photoelectron as the final state, such that the selection rules of bound-to-bound transitions hold. Defining the transition dipole with an orbital basis, the intensity of a transition in an electron detachment event can be determined by considering the integral
\begin{align}
I\propto&
\left\{
\langle \Psi^{N-1}(\Lambda_f)|
\otimes
\langle \psi_{\mathbf{k}}(l,\lambda_l)|
\right\}
\hat{\mu}
\left\{
|\Psi^N(\Lambda_i)\rangle\otimes\vert\rangle
\right\}
    \label{eq:intensity}
\end{align}
where $\vert\Psi^{N}(\Lambda_{i})\rangle$ is the initial molecular wavefunction with angular momentum projected on the internuclear axis of $\Lambda_{i}$, $\vert\Psi^{N-1}(\Lambda_{f})\rangle$ is the final molecular wavefunction with angular momentum projected on the internuclear axis of $\Lambda_{f}$, $\vert\psi_{\mathbf{k}}(l,\lambda_{l})\rangle$ is the photoelectron angular momentum with angular momentum quantum number $l$ and projected angular momentum quantum number $\lambda_{l}$, $\mathbf{k}$ is the photoelectron momentum vector and $\hat{\mu}$ is the dipole operator. For simplicity, we write both states assuming Hund's case (a) and represent orbital angular momentum only,\cite{Xie.1990} but where spin orbit coupling is strong, the equivalent Hund's case (c) is assumed, with the respective $\Omega$ and $\omega$ quantum numbers substituted into eq.\ \ref{eq:intensity}. 

For an electric-dipole transition within a spin-free closed system such as that represented by eq.\ \ref{eq:intensity}, the diatomic selection rule is
\begin{equation}
    \Delta\Lambda_{\text{total}} = 0,\pm1,\text{ and }\Delta S_{\text{total}}=0,
\end{equation}
where the change in total system orbital angular momentum $\Lambda_{\text{total}}$ arises from the projection of the dipole operator on the molecular axis and $S_{\text{total}}$ is the total system spin angular momentum. Therefore, the orbital angular momentum selection rule is
\begin{equation}
\Lambda_{f}+\lambda_{l} = \Lambda_{i} + \{0,\pm1\}\Rightarrow \{\lambda_{l}\} =  \{ - \Delta \Lambda+q|q\in\{0,\pm1\}\},
\label{eq:orbital_selection_rule}
\end{equation}
while the spin angular momentum selection rule, given that the photoelectron spin angular momentum is $s=\pm\frac{1}{2}$, is $\Delta S=S_{f}-S_{i}=\pm\frac{1}{2}$. When spin-orbit coupling is strong, spin and orbital angular momentum are no longer good quantum numbers, and the appropriate quantum number is $\Omega$. Given the projected spin selection rule $\Sigma_{f}+\sigma=\Sigma_{i}$, where $\sigma$ and $\Sigma_{\{i,f\}}$ are the photoelectron and molecular spin angular momentum projected onto the internuclear axis respectively, the selection rule in eq.\ \ref{eq:orbital_selection_rule} is modified to 
\begin{equation} \label{Angular momentum}
    \Omega_f + \omega_{j} = \Omega_i + \{0,\pm1\},
\end{equation} 
where $\omega_{j}$ is the total projected photoelectron angular momentum which can take values of $\omega_{j}=\vert\lambda_{l}-\sigma\vert,\ldots,\lambda_{l}+\sigma$. Similarly, $\Omega_{\{i,f\}}=\vert\Lambda_{\{i,f\}}-\Sigma_{\{i,f\}}\vert,\ldots,\Lambda_{\{i,f\}}+\Sigma_{\{i,f\}}$. 

The projected photoelectron angular momentum $\lambda_{l}$ is related to $l$ as 
\begin{equation}
    \lambda_{l} = -l,\ldots, +l \Rightarrow l\ge \vert \lambda_{l} \vert.
\end{equation}
The allowed $l$ define the nonzero terms in the photoelectron partial wave expansion,
\begin{equation}
\vert\psi_{\mathbf{k}}\rangle = \sum_{l=0}^{\infty}\sum_{\lambda_{l}=-l}^{+l}a_{l\lambda_{l}}R_{kl}(r)Y_{l\lambda_{l}}(\theta,\phi),    
\end{equation}
in which $a_{l\lambda_{l}}$ is the expansion coefficient, $Y_{l\lambda_{l}}(\theta,\phi)$ is the spherical harmonic describing the angular part of the wavefunction, and $R_{kl}$ is the radial wavefunction. For strong spin-orbit coupling, the total angular momentum is used and the possible $j$ in the partial wave expansion are $j\ge\vert\omega_{j}\vert$. It should be noted that spin-orbit coupling does not lead to significant energy splitting for the photoelectron, but is required for enforcing angular momentum conservation of the total final system.\cite{RevModPhys.54.389} 

Having described the state selection rules for the photoelectron partial wave expansion, we now examine the selection rules for the orbital from which photodetachment occurs $\vert\phi_{p}\rangle$. Expanding eq.\ \ref{eq:intensity} gives
\begin{equation} 
I\propto\sum_p
\langle \Psi^{N-1}(\Lambda_f)|a_p|\Psi^N(\Lambda_i)\rangle
\langle\psi_{\mathbf{k}}(l,\lambda_l)|\hat{\mu}|\phi_p(\lambda_{\ell})\rangle
+
\langle \Psi^{N-1}(\Lambda_f)|\hat{\mu}a_p|\Psi^N(\Lambda_i)\rangle
\langle\psi_{\mathbf{k}}(l,\lambda_l)|\phi_p(\lambda_{\ell})\rangle,\label{eq:intensity2}
\end{equation}
where $a^{\dagger}_{p}$ creates an electron in the $p^{th}$ orbital and $\lambda_{\ell}$ is the detachment orbital angular momenta projected onto the interatomic axis. As for eq.\ \ref{eq:intensity}, eq.\ \ref{eq:intensity2} can be written using $\Omega$ and $\omega$ quantum numbers when spin orbit-coupling is strong. The first term in eq.\ \ref{eq:intensity2} corresponds to direct photodetachment, in which the electron is ejected by photon excitation from a bound to a continuum state. The second integral in this first term results in the photodetachment channel selection rule $\lambda_{\ell}\in\{\lambda_{l}+q|q\in\{0,\pm1\}\}$. The first integral in this first term is the Dyson amplitude $\gamma_{p}$, which gives the contribution of orbital $\vert\phi_{p}\rangle$ to the Dyson orbital $\vert\phi_{D}\rangle$, 
\begin{equation}
    \vert\phi_{D}\rangle = \sqrt{N}\int{\Psi_{f}^{N-1}}^{*}(\mathbf{x}_{2},\ldots,\mathbf{x}_{N}){\Psi_{i}^{N}}(\mathbf{x}_{1},\mathbf{x}_{2},\ldots,\mathbf{x}_{N})d\mathbf{x}_{2}\ldots d\mathbf{x}_{N}, \label{eq:Dyson}
\end{equation}
such that the norm of the Dyson orbital is $\|\vert\phi_{D}\rangle\|^{2}=\sum_{p}\vert\gamma_{p}\vert^{2}$. In this integral, the annihilation operator removes the orbital with projected angular momentum $\lambda_{\ell}$ from $\langle\Psi^{N}\vert$ so that integral is only non-zero when $\lambda_{\ell} = -\Delta\Lambda$. As $\|\vert\phi_{D}\rangle\|^{2}$ determines the intensity of the direct process, one-electron ejection processes with no or little orbital relaxation result in an intense peak in the spectrum. However, where there is significant orbital relaxation in response to photoionization, $\|\vert\phi_{D}\rangle\|^{2}$ is substantially reduced and the signal is much less intense. Additionally, as the initial and final states are generally expressed as a multiconfigurational expansion, it is possible that shake-up processes involving valence `excitations' that result in zero $\gamma_{p}$ for photodetachment from the leading configuration can have non-zero components from minor configurations. In these cases, the intensity of the signal will be scaled proportionally to the contribution of these minor configurations to the total wavefunction.\cite{Amusia1975,brundle1979electron}  

The second term describes correlated two-electron processes in which the photon induces a bound-to-bound electronic transition while a second electron is emitted through a monopole transition. Such processes are referred to as conjugate shake-up transitions and behave as a concerted excitation--ionization event.\cite{starace1982theory,thomas1984transition,ungier1984resonance,schirmer1991satellite,fujiwara2005excitation} The initial and final angular momentum in the term are related as $\Lambda_i-\lambda_{\ell}=\Lambda_{f}+\{0,\pm1\}$, leading to a selection rule of $\lambda_{\ell}\in\{\Delta\Lambda+q|q\in\{0,\pm1\}\}$. The second integral selects for $\lambda_{\ell}\in\{\lambda_{l}\}$, which owing to the definition of $\{\lambda_{l}\}$ is the same selection rule as the first integral. 

The two terms in eq.\ \ref{eq:intensity2} distinguish two classes of photodetachment amplitude according to how the photon angular momentum enters the many-electron transition. However, while these classifications specify the photon interaction mechanism, they do not specify the dynamics of electron detachment. Both direct and conjugate processes can be modified by the presence of resonances or \PEVE{} interactions.\cite{Bilodeau.2012}

Upon ionization the photoelectron can experience a potential barrier capable of temporarily confining the electron. Such an effect is known as a shape resonance and can result in an energy-dependent enhancement of the photoionization cross section and potentially large changes in the photoelectron angular distribution. Consequently, while the orbital and state selection rules do not change, the intensity can be redistributed into the allowed final electronic states that contain contribution from photoelectron partial waves that are enhanced by the shape resonance.

An alternative resonance mechanism arises due to the presence of quasibound electronic states that lie above the detachment threshold, known as a Feshbach resonance. Excitation to these metastable states leads to ionization lifetimes determined by the associated decay width. As the Feshbach resonance first involves an excitation, the process is governed by the bound-to-bound electric-dipole selection rule between the initial and resonant anion states. The subsequent autodetachment is then determined by coupling of the resonance to the continuum. If the electron that is excited is later detached the resonance exhibits participator decay, while if a different electron is ejected to that which is excited, the resonance exhibits spectator decay. As a result, a final state that is weakly accessible through the direct one-electron Dyson amplitude can acquire appreciable intensity if it is strongly coupled to the resonance that is populated. In addition to the selection rules that govern the overall process (direct or conjugate shake-up), the bound-to-bound selection rule must allow the metastable state to be populated.

An alternative mechanism that allows neutral states to acquire intensity without formation of a formal metastable resonance state is through \PEVE{} interactions. In this mechanism, a time-dependent potential arising from changing electron-electron and electron-nuclear interactions during the photodetachment event results in nonadiabatic transitions in the remnant molecule. This mechanism has been proposed based on experimental observations,\cite{mason2021photoelectron,Mason2021,Huizenga.202555o,Huizenga.2025} and becomes increasingly important for low-kinetic-energy electrons as indicated by time-dependent simulations.\cite{Kinyua.2025} The effect of \PEVE{} interactions can be observed through inserting the resolution of the identity $\sum_{q}\vert\phi_{q}\rangle\langle\phi_{q}\vert=1$ into the direct term in eq.\ \ref{eq:intensity2} as a concerted mechanism, 
\begin{equation} 
I_{\text{direct}}\propto
\sum_{pq}
\mu_{pq}
\langle\Psi^{N-1}(\Lambda_f)|a_p|\Psi^{N}(\Lambda_i)\rangle
\langle\psi_{\mathbf{k}}(l,\lambda_l)|\phi_q(\lambda_{\ell}^{q})\rangle,\label{eq:intensity3}
\end{equation}
in which the photoelectron is coupled with the remnant molecule through the transition dipole moment $\mu_{pq}$.
The first integral in this term stipulates that $\lambda_{\ell}^{p}=-\Delta\Lambda$, where $\lambda_{\ell}^{p}$ is the projected orbital angular momentum of the orbital connecting initial and final states. However, the second integral is non-zero if $\lambda_{\ell}^{q}=\lambda_{l}$. In addition, $\lambda_{\ell}^{p}$ must be equal to $\lambda_{\ell}^{q}+\{0,\pm1\}$ from the angular momentum conservation requirements arising from the dipole operator. Therefore, substituting the relationship between the projected angular momentum of the different orbitals into the first selection rule, we can show that $\lambda_{\ell}^{q}=-\Delta\Lambda+\{0\pm1\}$ which is the conjugate shake-up selection rule. Therefore, \PEVE{} interactions can be seen to mix direct and conjugate-shakeup mechanisms, which results from the lack of a clear separation between the photoelectron and the bound electrons.

\subsubsection{Effects of angular distribution and shake-up transitions on intensity}
As discussed in Section \ref{ssec:experimental}, the intensity also has an angular distribution that is dependent upon the $\beta$ parameter through eq.\ \ref{eq:beta}. The anisotropy parameter depends on the photoelectron partial wave interference according to the Cooper-Zare equation\cite{cooper1968angular}
\begin{equation}
 \beta\approx\frac{2\Re(D_{l+1}D^{*}_{l-1})}{\vert D_{l+1}\vert^{2}+\vert D_{l-1}\vert^{2}},
\end{equation}
where $D_{l}$ is the transition dipole matrix element between the initial bound state and final continuum state. For direct ionization processes (first term in eq.\ \ref{eq:intensity2}), $D_{l}$ is the transition dipole moment between the origin orbital with angular momentum $\ell$ and the final continuum state with partial wave $l$. Therefore, for detachment from an s orbital ($\ell=0$), only a p-wave continuum channel exists ($l+1=1$), typically yielding strongly anisotropic emission with $\beta \rightarrow 2$ near threshold. In contrast, detachment from $p$, $d$ or $f$ orbitals introduces multiple outgoing partial waves whose interference reduces directional coherence, resulting in smaller $\beta$ values and more isotropic emission. Thus, systematic changes in $\beta$ directly reflect changes in the orbital angular momentum composition of the Dyson orbital associated with each transition.

In heavy-element systems, SOC further modifies the observable angular distributions by mixing states of different spin and spatial symmetry prior to detachment.\cite{mishra2006importance} Consequently, the measured $\beta$ parameter reflects not only the parent orbital angular momentum but also SOC-mediated configuration mixing, particularly in regions of high state density. 

For conjugate shake-up transitions (second term in eq.\ \ref{eq:intensity2}), the photoionization involves a correlated two-electron process so that the Cooper-Zare model can no longer be strictly applied. Therefore, the photoelectron angular distribution is modified by the orbitals involved in the bound-to-bound transition such that it can deviate markedly from the expected one-electron detachment. 

\subsection{Application to CeO}
\begin{figure}[!ht]
\centering
\includegraphics[width=16cm]{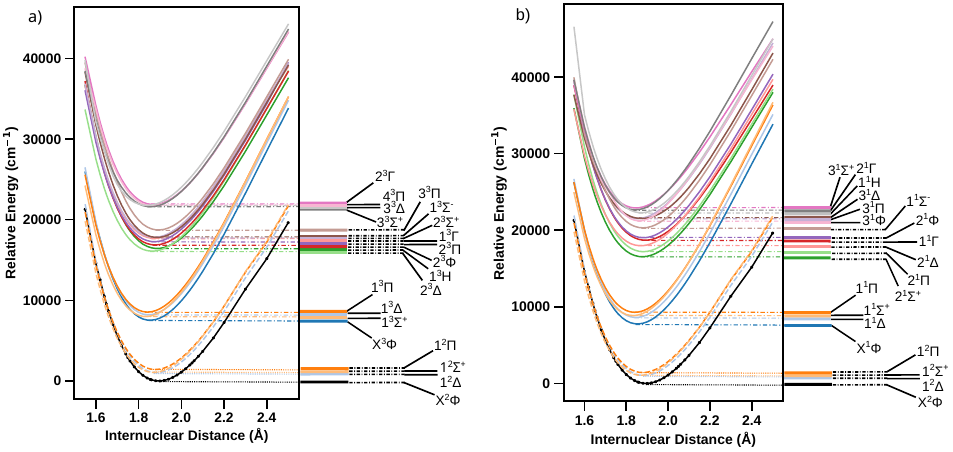}
 \caption{Potential energy curves of the ground and selected low-lying excited states of CeO$^{-}$ together with the low-lying triplet (a) and singlet (b) states of neutral CeO along the Ce--O bond coordinate. The black curve denotes the anion ground state, while the colored dotted and solid curves represent excited anion and neutral states, respectively. Markers on the black curve indicate the calculated energy points used to construct the potential energy curve. Energies are given in cm$^{-1}$ relative to the minimum of the anion ground state.}
    \label{fig:pec_spin_free}
\end{figure}

Comparison of CASSCF and NEVPT2 detachment energies (Table~S3) demonstrates the importance of dynamic correlation, with NEVPT2 providing excellent agreement with experiment.\cite{cantero2021theoretical} On this basis, Figure~\ref{fig:pec_spin_free} presents the spin-free CASSCF+NEVPT2 potential energy curves of the low-lying states of CeO and \ce{CeO^-}.

 For \ce{CeO^-}, only the low-lying doublet states below the neutral detachment threshold are shown, as the quartet states are energetically inaccessible. In line with previous studies,\cite{ray2015photoelectron,makhlouf2021theoretical} both the anion and neutral ground states have $\Phi$ symmetry and their lowest electronic manifolds can be described as different permutations of a single $4f$ electron. 
The similarity between the anion and neutral species also extends to the second electronic manifold, which can be qualitatively described as a $\sigma \rightarrow 5d$ excitation lying approximately 10,000 cm$^{-1}$ above the respective ground states. However, near the equilibrium geometry, these excited anion states lie above the neutral ground state and are therefore metastable with respect to autodetachment. Consequently, the corresponding excited rovibronic 
states become non-stationary and cannot be quantitatively described within the conventional NEVPT2 framework. 

The near-degeneracy between the lowest set of singlet and triplet states  of the neutral system confirms the expected small non-relativistic coupling between the $4f^1$-core and the valence $\sigma$ open-shell electron, highlighting the essential role of SOC in determining the low-lying electronic structure (see below). The calculated low-lying neutral states up to \(\approx 22{,}000\) cm\(^{-1}\) arise predominantly from different occupations and couplings of the valence orbitals. They have significant contributions from the Ce-centered \(4f\) and \(5d\) atomic orbitals and correspond mainly to the \(6s^14f^1\), \(5d^2\), and \(4f^15d^1\) configuration manifolds. The spin-free PECs reveal two distinct electronic configurational regimes. The lower-lying manifold is primarily associated with $6s^{1}5d^{1}$ character and exhibits equilibrium bond lengths shifted relative to the anion ground state, indicating a change in bonding upon electron detachment. In contrast, the higher-lying states, dominated by $4f^{1}5d^{1}$ character, display minima located near the anion equilibrium geometry, consistent with the more localized nature of the $4f$ orbital. 
 
 The spectroscopic constants corresponding to the spin-free PECs of anionic and neutral CeO are listed in Table S5, while those for the spin-orbit coupled states of neutral CeO are given in Table \ref{tab:soc_results}. 
These spectroscopic constants, combined with the calculated PECs (Figure~\ref{fig:pec_spin_free}) and CASSCF orbitals (Figure S3), show that the Ce-O bond strength is governed primarily by the extent of Ce $5d$--O $2p$ covalent interaction. In the CeO ground state, the \(5d\)-derived \(\sigma(d_{z^2})\), \(\pi(d_{xz},d_{yz})\), and \(\delta(d_{xy},d_{x^2-y^2})\) orbitals exhibit significant polarization toward the O center, consistent with enhanced covalency and shorter equilibrium bond length (\(R_e \approx 1.820\)~\AA). In contrast, the Ce \(4f\) orbitals remain largely localized on CeO center with minimal O \(2p\) overlap, indicating predominantly nonbonding character. The diffuse appearance of \(d_{xz}/d_{yz}\) and \(d_{z^{2}}\) orbitals in Figure S3 reflects their broader spatial extent. The longer equilibrium bond length of the anion (\(R_e \approx 1.855\)~\AA) indicates reduced Ce--O covalency upon electron attachment. Enhanced localization of the density to Ce center and reduced \(5d\)--\(2p\) overlap weaken the anionic bond relative to neutral, leading to a longer equilibrium distance. Similar trends are observed for the excited-state PECs, where states with greater $5d$ character exhibit shorter equilibrium distances, whereas \(4f\)-dominated states retain geometries closer to that of the anion owing to the localized, weakly bonding nature of the \(4f\) orbitals. Mixed configurations give rise to intermediate equilibrium distances and enhanced Franck--Condon overlap with the anion ground state.
\begin{figure}[!ht]
\centering
\includegraphics[width=14cm]{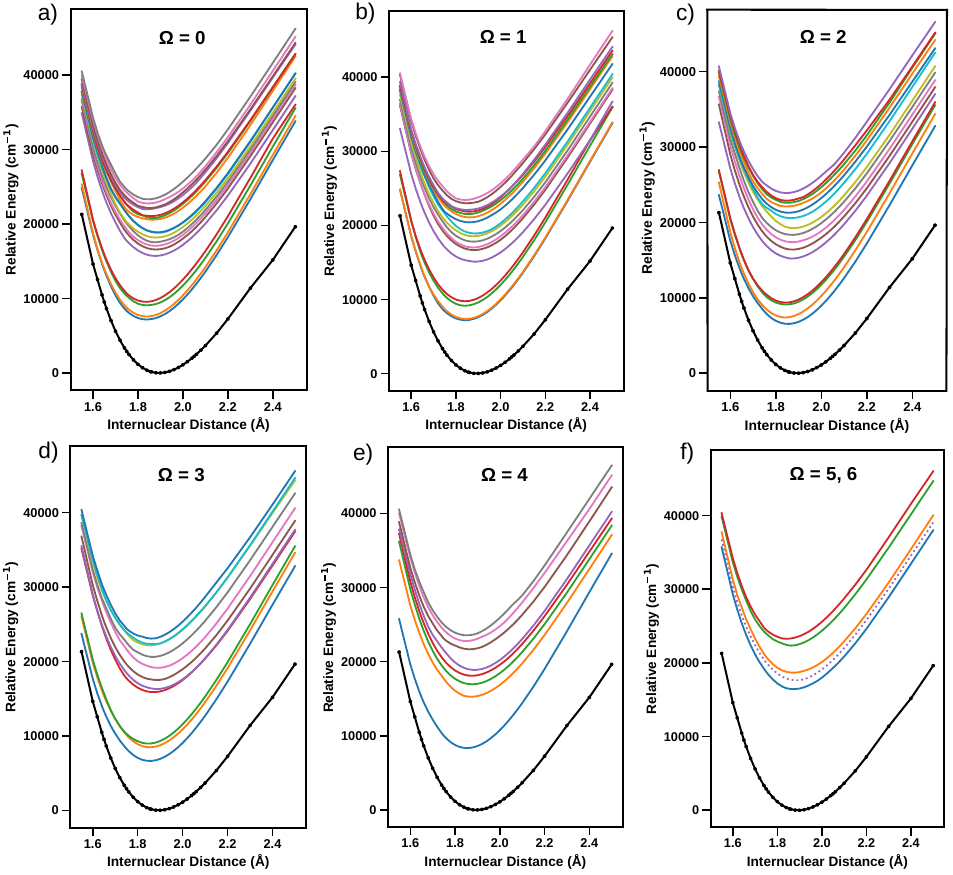}
\caption{Spin-orbit-coupled potential energy curves of CeO grouped by the $\Omega$ quantum number: (a) $\Omega = 0$, (b) $\Omega = 1$, (c) $\Omega = 2$, (d) $\Omega = 3$, (e) $\Omega = 4$, and (f) $\Omega = 5$ (solid line) and $\Omega = 6$ (dotted line). The black curve denotes the CeO$^{-}$ ground state, with markers indicating calculated points. Energies are given in cm$^{-1}$ relative to the minimum of the anion ground state. } 
    \label{fig:pec_soc}
\end{figure}

Figure \ref{fig:pec_soc} shows the spin-orbit coupled NEVPT2 PECs of CeO resolved into individual $\Omega$ components ($\Omega$ = 0-6). All energies are given relative to the minimum of the anion ground-state potential energy curve, shown as the black curve. Within each panel, the spin-orbit interaction lifts the degeneracy of the parent spin-free states and generates a set of closely spaced curves that retain similar overall shapes but differ in their equilibrium positions and relative ordering (Section S3 in SI). The lowest-energy curve occurs in the $\Omega = 2$ manifold and defines the ground state, consistent with its dominant $4f^{1}\sigma^{1}$ parentage and with previous experimental and theoretical results.\cite{ray2015photoelectron,makhlouf2021theoretical} Across the $\Omega = 0$-3 panels, the high density of low-lying states leads to pronounced mixing and several avoided crossings, reflecting strong interaction among near-degenerate components of the $6s^{1}4f^{1}$ manifold. In contrast, the higher $\Omega$ manifolds ($\Omega$ = 4, 5) display a more regular and parallel set of curves with reduced mixing and fewer avoided crossings, primarily reflecting the smaller number of electronic states contributing to the high-$\Omega$ manifolds and the resulting lower state density. Comparison with the spin-free representation shows that SOC primarily redistributes the states among the different $\Omega$ components and modifies their detailed ordering, while preserving the overall separation into lower $6s^{1}4f^1$, $5d^2$ 
and higher $4f^{1}5d^{1}$ manifolds. Consequently, the photodetachment intensity pattern remains governed by the parent configuration character, with SOC introducing a fine-structure splitting of the spectral features across the $\Omega$ components. 

In our CASSCF+NEVPT2 results, the $\Omega = 0$ and 1 states with dominant $^3\Sigma^+$ character appear very close in energy to the X$^3\Phi_2$ ground state. This unusually strong stabilization of the $^3\Sigma^+$ parent term is at odds with the experimental picture, in which the lowest manifold is consistently assigned as predominantly $^3\Phi$, with the low-lying excited states arising from $^3\Delta$, $^3\Pi$, and $^3\Sigma^+$ parent terms.\cite{linton1983electronic,kaledin1993laser} Given the known sensitivity of multireference perturbation theory excitation energies to basis-set size, active-space choice, and differential dynamical correlation between states of different character, we consider this low-lying $^3\Sigma^+$ to be at least partly a computational artefact.\cite{angeli2001n, lischka2018multireference} A more balanced description (for example including additional $6p$ and the oxygen valence orbitals in the active space and improving the treatment of core-valence correlation) is expected to  increase the relative energy gap between $^3\Sigma_{0,1}^+$ states and the ground state. 

\begin{figure}[!t]
\centering
\includegraphics{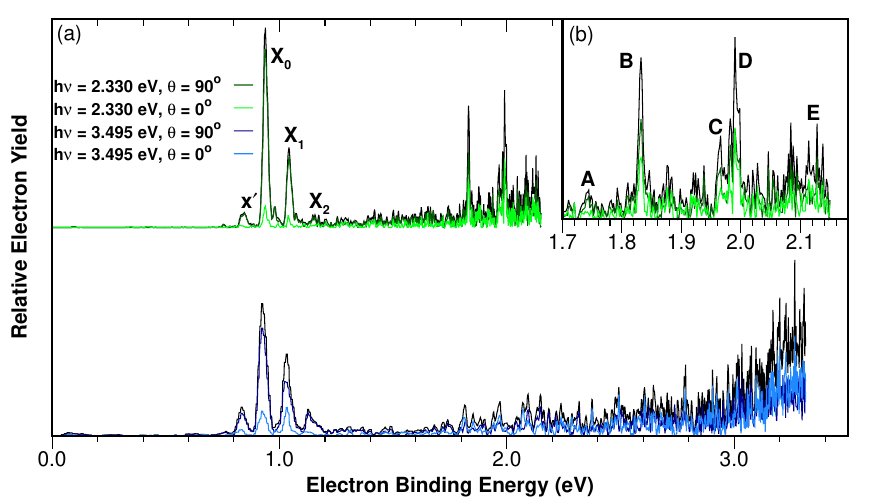}
 \caption{(a) High-resolution anion photoelectron spectra of \ce{CeO^-} recorded at photon energies of 2.330 and 3.495 eV with the laser polarization aligned parallel (light green and light blue) and perpendicular (dark green and dark blue) to the electron detection axis. The black line shows the sum of parallel and perpendicular spectra. (b) Expanded view of the 1.70--2.18 eV binding energy region, highlighting the spectral features used for the electronic-state assignments. The labeled peaks A-E correspond to the transitions analysed in this work.}
   \label{fig:exp}
\end{figure}

\renewcommand{\arraystretch}{0.7}
\begin{longtable}{ccccc| cccc}
\caption{Computed spin-orbit coupled spectroscopic constants for CeO at the CASSCF+NEVPT2 level of theory. Energies are given relative to the lowest $\Omega$ state. In the state label $m\,\Omega=n$, $n$ denotes the $\Omega$ value and $m$ the energetic ordering of states within a given $\Omega$ manifold. $T_e$, $R_e$, and $\omega_e$ represent the electronic excitation energy (cm$^{-1}$), equilibrium bond distance (\AA), and harmonic vibrational constant (cm$^{-1}$), respectively. Assign. denotes the spectral assignment based on the experimental spectrum shown in Figure~\ref{fig:exp}.}
\label{tab:soc_results} \\
\hline
State & $T_e$ & $R_e$ (\AA) & $\omega_e$ &  Assign. &State & $T_e$ &  $R_e$ (\AA)& $\omega_e$   \\
\hline
\endfirsthead

\hline
State & $T_e$  &  $R_e$  & $\omega_e$  &  Assign. &State & $T_e$  &  $R_e$  & $\omega_e$  \\
\hline
\endhead
X$\Omega$=2 & 0     & 1.8532 & 811 & & 9 $\Omega$=1 & 11996 & 1.8789 & 741  \\
1 $\Omega$=3 & 94    & 1.8531 & 795 & & 2 $\Omega$=5 & 12140 & 1.8721 & 724 \\
1 $\Omega$=0 & 642   & 1.8376 & 833 & & 10 $\Omega$=0 & 12285 & 1.8879 & 765\\
1 $\Omega$=1 & 668   & 1.8425 & 765 & &11 $\Omega$=0 & 12349 & 1.8868 & 744 \\
2 $\Omega$=1 & 827   & 1.8456 & 798 & &6 $\Omega$=4 & 12349 & 1.8868 & 718 \\
2 $\Omega$=2 & 849   & 1.8411 & 842 & & 10 $\Omega$=1 & 12374 & 1.8868 & 751\\
2 $\Omega$=0& 1043  & 1.8392 & 817 &  & 7 $\Omega$=3 & 12593 & 1.8891 & 788 \\
1 $\Omega$=4 & 1809  & 1.8536 & 802 & & 9 $\Omega$=2 & 12711 & 1.8780 & 743 \\
2 $\Omega$=3 & 1929  & 1.8535 & 788 & &  11 $\Omega$=1 & 13871 & 1.8609 & 721   \\
3 $\Omega$=3 & 2425  & 1.8472 & 795 & & 12 $\Omega$=0 & 14044 & 1.8652 & 748  \\
3 $\Omega$=0 & 2547  & 1.8393 & 801 & &10 $\Omega$=2 & 14045 & 1.8646 & 639   \\
3 $\Omega$=2 & 2571  & 1.8462 & 856 & &8 $\Omega$=3 & 14060 & 1.8683 & 802  \\
3 $\Omega$=1 & 2613  & 1.8394 & 756 & &13 $\Omega$=0 & 14246 & 1.8669 & 868   \\
4 $\Omega$=2 & 2826  & 1.8437 & 837 & & 14 $\Omega$=0 & 14538 & 1.8632 & 858   \\
4 $\Omega$=0 & 3013  & 1.8360 & 815 & &12 $\Omega$=1  & 14540 & 1.8618 & 741  \\
4 $\Omega$=1 & 3223  & 1.8411 & 805 & &11 $\Omega$=2 & 14746 & 1.8566 & 764  \\
5 $\Omega$=1 & 8563  & 1.8876 & 792&B &13 $\Omega$=1 & 14986 & 1.8557 & 746  \\
5 $\Omega$=2 & 8668  & 1.8686 & 929&B & 7 $\Omega$=4 & 15138 & 1.8656 & 739  \\
2 $\Omega$=4 & 8738  & 1.8721 & 676&B & 14 $\Omega$=1 & 15308 & 1.8576 & 710  \\
5 $\Omega$=0 & 9179  & 1.8767 &709 &  & 15 $\Omega$=0 & 15440 & 1.8384 & 593  \\
4 $\Omega$=3 & 9337 & 1.8699 &742&C/D & 15 $\Omega$=1 & 15544 & 1.8474 & 777  \\
5 $\Omega$=3 & 9754 & 1.8855 &727&C/D & 12 $\Omega$=2 & 15583 & 1.8470 & 472  \\
6 $\Omega$=2 & 9857& 1.8725 &739 &C/D & 16 $\Omega$=0 & 15602 & 1.8552 & 942  \\
1 $\Omega$=5 & 9910  & 1.8726 & 741 & & 9 $\Omega$=3 & 15602 & 1.8554 & 948  \\
6 $\Omega$=0 & 10039 & 1.8803 & 743 & &10 $\Omega$=3 & 15795 & 1.8749 & 631  \\
6 $\Omega$=1 & 10134 & 1.8813 & 734 & &3 $\Omega$=5 & 15806 & 1.8620 & 1081  \\
3 $\Omega$=4 & 10417 & 1.8766 & 744 & &13 $\Omega$=2 & 16070 & 1.8531 & 703  \\
7 $\Omega$=1 &10482& 1.8765& 737 &   &17 $\Omega$=0 & 16222 & 1.8464 & 780  \\
7 $\Omega$=0 & 10536 & 1.8782 & 728 & & 8 $\Omega$=4 & 16250 & 1.8497 & 788 \\
7 $\Omega$=2 & 10847 & 1.8721 & 689&E &14 $\Omega$=2 & 16343 & 1.8489 & 785  \\
6 $\Omega$=3 & 10959 & 1.8854 & 838&E & 16 $\Omega$=1 & 16454 & 1.8480 & 723 \\
8 $\Omega$=0 & 11013 & 1.8720 & 856& & 11 $\Omega$=3 & 16525 & 1.8638 & 838  \\
1 $\Omega$=6 & 11093 & 1.8855 & 756 & &4 $\Omega$=5 & 16738 & 1.8384 & 651  \\
8 $\Omega$=1 & 11267 & 1.8792 & 745& &18 $\Omega$=0 & 16782 & 1.8439 & 613  \\
5 $\Omega$=4 & 11564 & 1.8757 & 723& &17 $\Omega$=1 & 16895 & 1.8423 & 809  \\
9 $\Omega$=0 & 11648 & 1.8809 & 722 & &9 $\Omega$=4 & 17024 & 1.8530 & 728  \\
8 $\Omega$=2 & 11812 & 1.8714 & 740 & &15 $\Omega$=2 & 17365 & 1.8467 & 772  \\
\hline
\end{longtable}
The experimental photoelectron  spectrum measured with 2.330 eV (green) and 3.495 eV (blue) photon energies, shown in Figure \ref{fig:exp}, exhibits an intense origin band X$_0$ at \(0.936 \pm 0.008\) eV, a weak low-binding-energy feature x$'$, and a vibrational progression X$_1$--X$_2$ at higher binding energies\cite{ray2015photoelectron}. The \ce{CeO^-} photoelectron spectrum is interpreted in terms of detachment from the X${}^{2}\Phi_{5/2}$ anion ground state, dominated by a \(6s^24f^1 \) configuration, to a low-lying SOC-split neutral \(^{3}\Phi\) manifold of predominantly \( 6s^14f^1\) character. Consequently, photodetachment is dominated by Ce \(6s\) electron removal, consistent with diffuse \(s\)-type Dyson orbitals in the low-binding-energy region. The X$_0$ band is assigned to the \(v'=0 \rightarrow v''=0\) transition into the neutral X$^{3}\Phi_{2}$ ground SOC state computed at \(0.94~\mathrm{eV}\). The X$_1$ and X$_2$ features are assigned to the corresponding vibrational progression in the neutral Ce-O stretching mode, while the weak x$'$ feature is attributed to a hot-band transition originating from vibrationally excited anion levels, \(v'=1 \rightarrow v''=0\). The calculated detachment energies and vibrational spacings show excellent agreement with experiment (Table~\ref{tab:assign}).

 Binning the photoelectron data into uniform electron binding energy intervals enhances the relative intensity of several weak features at higher electron binding energies. Interpretation of these features requires a reliable description of the low-lying electronic structure of both the anion and the neutral. We therefore begin by examining the low-lying spin-orbit-coupled (SOC) states of CeO$^{-}$ computed at the NEVPT2 level before considering the corresponding neutral states and the observed photodetachment transitions. Within the energy window close to the experimental photon energy between 2.00 and 2.33 eV, six optically accessible SOC states are identified as potential intermediate resonances in the photodetachment process (Table S7). Although direct photodetachment from the anion ground state is energetically allowed, substantial electronic relaxation is required following detachment from the dominant \(6s^24f^1\) configuration. This relaxation is expected to reduce the direct one-electron contribution. In contrast, excitation to the bright anion resonance provides a more favorable electronic pathway for subsequent electron detachment. A quantitative comparison of the direct and resonance-mediated photodetachment pathways would require explicit evaluation of the corresponding Dyson orbitals and photoionization matrix elements.\cite{oana2009cross,tenorio2022photoionization} Accordingly, the present interpretation is based on the electronic configurations of the initial and final states. Four of these states (at 2.11, 2.13, 2.15, and 2.15 eV) are dominated by the $|\sigma_s\pi_d\phi_f\rangle$ configuration, with smaller contributions from $|\sigma_s\sigma_d\pi_f\rangle$ and $|\sigma_s\sigma_d\delta_f\rangle$. Analysis of the neutral SOC wavefunctions accessible through direct ionization of these configurations indicates that these anion states are unable to serve as candidates for the observed autoionizing resonance. In contrast, the bright SOC states at 2.05 and 2.12 eV show pronounced configurational mixing. They contain substantial contributions from configurations that also contribute significantly to the relevant neutral SOC states upon ionization, making them the most plausible candidates for the observed resonance.
 Higher-lying states between 2.15 and 2.33 eV exhibit substantially weaker oscillator strengths despite their closer energetic proximity to the experimental photon energy and so were not considered further as potential resonance states.

The lower-energy candidate resonance at 2.05~eV exhibits significant multiconfigurational character, with its wavefunction dominated by contributions from the spin-free $^{2}\Sigma$, $^{2}\Pi$, and $^{2}\Phi$ manifolds. The leading configuration-state functions are $|\sigma_s\delta_d\delta_f\rangle$ ($^{2}\Sigma$, 22\%), $|\sigma_s\delta_d\pi_f\rangle$ ($^{2}\Pi$, 16\%), $|\sigma_s\delta_d\phi_f\rangle$ ($^{2}\Pi$, 16\%), $|\sigma_d\delta_d\delta_f\rangle$ ($^{2}\Sigma$, 13\%), and $|\sigma_d\delta_d\pi_f\rangle$ ($^{2}\Phi$, 9\%) (Table S7). The higher-energy candidate resonance at 2.12~eV likewise displays strong multiconfigurational character, with dominant contributions from the same spin-free manifolds, together with a smaller contribution from the $|\sigma_s\pi_d\phi_f\rangle$ configuration of $^{4}\Delta/^{4}\Gamma$ character (10\%). As a result, these configurations couple to give states with $\Omega=\frac{1}{2}\ldots\frac{9}{2}$. Given the $\Omega = 5/2$ ground state of CeO$^{-}$, the subsequent analysis is restricted to the optically accessible resonance components with $\Omega = 3/2$, $5/2$, and $7/2$, in accordance with the electric-dipole selection rule,\cite{xie1990selection} $\Delta\Omega = 0,\pm1$.
\begin{table}[!htbp]
\centering
\caption{CASSCF+NEVPT2 assignments for the \ce{CeO^-} photoelectron spectrum. Experimental electron binding energies (BE, eV) and $\beta$ anisotropy parameters are taken from Figure \ref{fig:exp}. Computed $T_i^{\mathrm{calc.}}$ (= $T_e$ + I.E.) values correspond to adiabatic ionization energies of neutral \ce{CeO} relative to its ground SOC component.}
\label{tab:assign}
\begin{tabular}{lccccl}
\hline
Peak & BE (eV)$^{a}$ & $\beta$ & Assignment & $T_i^{\mathrm{calc.}}$ (eV) \\
\hline

x$'$  & 0.828 - 0.848        & 1.75 & ${X}^{3}\Phi_{2}$ ($v'=1 \rightarrow v''=0$)  & 0.84 \\

X$_0$ & $0.936 \pm 0.008$   & 1.75 & ${X}^{3}\Phi_{2}$ ($v'=0\rightarrow v''=0$)                    & 0.94 \\
X$_1$ & $1.042 \pm 0.010$   & 1.75 & ${X}^{3}\Phi_{2}$ ($v'=0\rightarrow v''=1$)                    & 1.04 \\
X$_2$ & $1.149 \pm 0.015$   & 1.75 & ${X}^{3}\Phi_{2}$ ($v'=0\rightarrow v''=2$)                    & 1.14 \\
A     & $1.74 \pm 0.01$     & 2.0  &  Tentative hot band                       & 1.89, 1.90, 1.91 \\

B    & $1.83 \pm 0.01$     & 0.3  & $5\Omega$=1 + $5\Omega$=2 + $2\Omega$=4  & 1.99, 2.00, 2.01 \\
C, D     & $1.96 \pm 0.01$, $1.99 \pm 0.01$      & 0.3  & $4\Omega$=3 + $5\Omega$=3 + $6\Omega$=2 & 2.09, 2.14, 2.15 \\
E     & $2.12 \pm 0.01$     & 0.3  & $7\Omega$=2 + $6\Omega$=3                               &  2.27, 2.29 \\

\hline
\end{tabular}

\begin{tablenotes}
\footnotesize
\item$^{a}$Experimental values from Ray et al. \textit{J. Chem. Phys.} \textbf{2015}, \textit{142}, 064305. 
\end{tablenotes}
\end{table}

Since these excited states lie above the electron-detachment threshold, they correspond to autoionizing resonances rather than bound electronic states. Consequently, their calculated excitation energies should be regarded as approximate, and a rigorous description of their resonance positions and lifetimes requires the inclusion of a complex absorbing potential (CAP).\cite{magoulas2015structural} Therefore, the present assignment is not based on the calculated resonance energies but rather on the electronic structure of the resonances, their oscillator strengths, and their ability to reproduce the experimental photoelectron intensity distribution.

The bright resonance is decomposed into its dominant electronic configurations, and one-electron detachment from the corresponding molecular orbitals is used to generate the accessible neutral electronic configurations. These configurations are subsequently matched with the calculated neutral SOC eigenstates according to their wavefunction compositions (Table S6). Relative transition intensities are estimated by weighting each photodetachment pathway using the contributions of the corresponding electronic configurations in both the excited anion and the neutral SOC states, and its initial oscillator strength, which is equivalent to
\begin{equation} 
I\propto\sum_I
\langle \Psi^{N-1}(\Lambda_f)|a_p|\Psi^N_I\rangle
\langle \Psi^N_I|\hat{\mu}|\Psi^N(\Lambda_i)\rangle
\langle\psi_{\mathbf{k}}(l,\lambda_l)|\phi_p(\lambda_{\ell})\rangle.
\end{equation}
Although this procedure does not provide quantitative photoionization cross sections, it yields a physically motivated qualitative description of the accessible photodetachment channels. 

\begin{figure}[!ht]
\centering
\includegraphics[width=12cm]{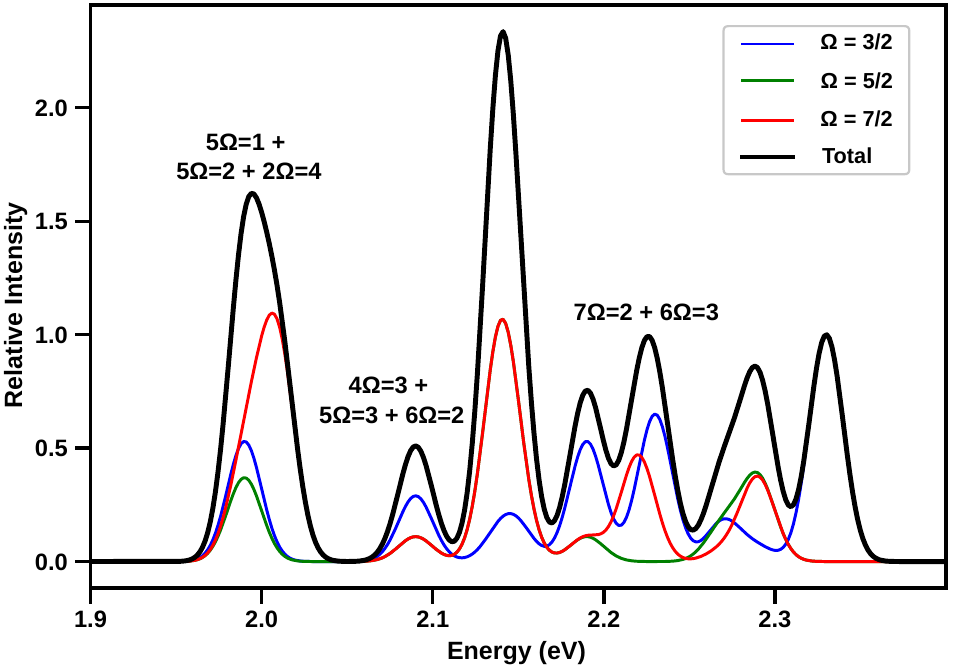}
\caption{Simulated photoelectron spectrum of CeO$^-$ obtained from the high-energy bright mixed spin--orbit coupled (SOC) resonance at 2.12~eV. The total spectrum is generated by applying Gaussian broadening (FWHM = 0.024~eV; $\sigma = 0.010$~eV) to the transitions to the neutral SOC states, weighted by the estimated relative photodetachment intensities derived from the wavefunction compositions. The individual contributions from the $\Omega=3/2$, $\Omega=5/2$, and $\Omega=7/2$ manifolds are also shown.}
   \label{fig:simulated_spectra}
\end{figure}

The simulated spectra generated from the two bright SOC resonances at 2.05 and 2.12~eV both reproduce the overall experimental band structure through unresolved groups of closely spaced SOC transitions (Figure \ref{fig:simulated_spectra} and Figure S6). However, the higher-energy resonance at 2.12~eV provides the more probable assignment of the experimental bands B--E, as it yields a more consistent distribution of spectral intensity and exhibits an oscillator strength (\(f = 4.23 \times 10^{-5}\)) approximately 2.3 times larger than that of the 2.05~eV resonance (\(f = 1.86 \times 10^{-5}\)), indicating a significantly higher transition probability. Given the small energy separation between the two resonances (0.07 eV), which is comparable to the expected uncertainty of the present calculations, contributions from both resonances cannot be excluded (Table S8 and S9). 

The weak low-binding-energy feature A is not reproduced by the present resonance simulations and therefore is tentatively assigned. Although its distinct photoelectron angular distribution may indicate a detachment mechanism different from that responsible for peaks~B--E, the limited signal-to-noise ratio precludes a definitive interpretation based on the measured anisotropy alone. The similarity between the experimental A--B energy separation ($0.09\pm0.01$~eV) and the x$'$--X$_0$ vibrational spacing ($\sim0.10$~eV) further suggests that peak~A may originate from a hot-band transition involving the same neutral states assigned to peak~B. 

The first intense experimental feature, peak~B, is assigned to a group of nearly degenerate neutral SOC states comprising $5\Omega=1$, $5\Omega=2$, and $2\Omega=4$ at 1.99--2.01~eV (Figure~\ref{fig:simulated_spectra}). The $5\Omega=1$ state provides the principal contribution across the calculated SOC manifolds, while the nearby $5\Omega=2$ and $2\Omega=4$ states further enhance the intensity through the $\Omega=7/2$ manifold. Consequently, peak~B arises from unresolved transitions to several closely spaced SOC states rather than from a single electronic transition. The calculated Dyson orbitals indicate that the dominant photodetachment channel involves removal of a $\delta$ electron, with smaller contributions from the $\sigma$ orbital. Therefore, this assignment is consistent with the selection rules presented in Section \ref{sec:selection_rules}. For the $\Omega=5/2$ ground state of CeO$^{-}$, the resonance-mediated selection rule, $\vert\omega\vert=\Delta\Omega+\{0,\pm1\}$, yields the allowed photoelectron projections $\omega=\{\pm1/2,\pm3/2,\pm5/2\}$ for the $5\Omega=1$ and $5\Omega=4$ states, and $\omega=\{\pm1/2,\pm3/2\}$ for the $5\Omega=2$ state. These correspond to photoelectron partial waves with orbital projections up to $|\lambda|=3$ for the $5\Omega=1$ and $2\Omega=4$ channels, and up to $|\lambda|=2$ for the $5\Omega=2$ channel, i.e.\ $d$ orbitals are the highest angular momentum photodetachment orbital possible from all channels. So, the experimentally observed anisotropy ($\beta\approx0.3$) is consistent with the predicted mixture of predominantly $d$-wave detachment.

The experimental intensity spanning peaks~C and~D is assigned to transitions involving the $4\Omega=3$, $5\Omega=3$, and nearby $6\Omega=2$ neutral SOC states. Peak~C is associated primarily with the $4\Omega=3$ state at 2.09~eV, whereas peak~D is dominated by the more intense $5\Omega=3$ transition at 2.14~eV, with additional contributions from the energetically adjacent $6\Omega=2$ state. Owing to the small energy separations between these SOC states and the experimental resolution, these transitions merge into a broad unresolved spectral envelope rather than appearing as distinct electronic features. The corresponding photodetachment channels are predominantly of $\sigma$ character, with smaller contributions from $\delta$ and mixed $\sigma/\delta$ configurations arising from SOC mixing. For the assigned $4\Omega=3$, $5\Omega=3$, and $6\Omega=2$ states, the resonance-mediated selection rules yield the allowed photoelectron projections $\omega=\{\pm1/2,\pm3/2\}$ correspond to orbital projections with $|\lambda|\le2$. Consequently, only up to $d$-wave photoelectron partial waves are symmetry allowed, in agreement with the predominantly $\sigma$- and $\delta$-orbital character obtained from the Dyson orbital analysis. 

The highest-binding-energy experimental feature (peak~E) is assigned to unresolved contributions from the closely spaced $7\Omega=2$ and $6\Omega=3$ neutral SOC states. This assignment is supported by the experimental D--E energy separation (0.13$\pm$0.02~eV), which is reproduced most closely by the calculated transitions at 2.27 and 2.29~eV than the higher-energy transition at 2.33~eV. The relative D--E energy separation serves only as a qualitative consistency check, the assignment is based primarily on the calculated intensity distribution and dominant orbital character. Although neither transition dominates the calculated intensity individually, their small energy separation is expected to yield a single unresolved experimental feature. The assigned states satisfy the same resonance-mediated selection rules as peaks~C and~D, permitting only $s$-, $p$-, and $d$-wave photoelectron partial waves. Their predominantly $\sigma$-orbital character, with smaller $\delta$ contributions arising from spin--orbit coupling, is therefore consistent with the symmetry-allowed detachment channels. The weak experimental intensity observed between peaks D and E is not reproduced by the present simulations and may reflect continuum photoionization or other dynamical effects beyond the present theoretical treatment. 

Overall, the calculated transition energies exhibit a systematic offset of approximately 0.1--0.2~eV relative to experiment. Such deviations are expected for bound-state descriptions of autoionizing resonances, for which the absolute resonance positions are generally more sensitive to the treatment of electron correlation and continuum coupling than the relative ordering and electronic character of the states.\cite{jagau2017extending} Accordingly, the spectral assignments presented here rely primarily on the relative ordering, calculated intensities, and electronic character of the computed states, with the absolute transition energies providing qualitative support.

\section {Conclusion}
This study combines high-resolution anion photoelectron spectroscopy with multireference relativistic electronic-structure calculations to investigate the electronic structure and photodetachment dynamics of \ce{CeO^-}. A binning-based spectral analysis has been used to enable observation of weak features within congested spectral manifolds, enabling more reliable interpretation of the photoelectron spectrum. Spin-free and spin--orbit-coupled CASSCF+NEVPT2 potential energy curves reproduce the low-lying electronic structure and detachment energetics in good agreement with experiment, providing a basis for tentative assignments of the observed photoelectron features. Analysis of anisotropy parameters and photodetachment selection rules further reveals the role of multielectron processes, including Feshbach resonances. The computed electronic manifolds exhibit substantial mixing among Ce-centered \(4f\), \(5d\), and higher-lying configurations, highlighting the importance of electron correlation and relativistic effects in the \ce{CeO^-}/CeO systems. Collectively, these results provide a detailed interpretation of the \ce{CeO^-} photoelectron spectrum  and establish a framework for investigating electronically complex lanthanide oxides. Future advances in theoretical treatments of orbital relaxation, multielectron excitations, and continuum-state interactions will further improve the quantitative description of lanthanide photodetachment phenomena.

\section{Acknowledgements}
This work was supported by the U.S. Department of Energy, Office of Science, Basic Energy Sciences, in the Gas Phase Chemical Physics and Computational and Theoretical Programs (grant no. DE-SC0024282). This study was performed using computational resources provided by the University of Louisville Research Computing Group and the Cardinal Research Cluster.

\bibliography{Reference}

\providecommand{\latin}[1]{#1}
\makeatletter
\providecommand{\doi}
  {\begingroup\let\do\@makeother\dospecials
  \catcode`\{=1 \catcode`\}=2 \doi@aux}
\providecommand{\doi@aux}[1]{\endgroup\texttt{#1}}
\makeatother
\providecommand*\mcitethebibliography{\thebibliography}
\csname @ifundefined\endcsname{endmcitethebibliography}  {\let\endmcitethebibliography\endthebibliography}{}
\begin{mcitethebibliography}{67}
\providecommand*\natexlab[1]{#1}
\providecommand*\mciteSetBstSublistMode[1]{}
\providecommand*\mciteSetBstMaxWidthForm[2]{}
\providecommand*\mciteBstWouldAddEndPuncttrue
  {\def\EndOfBibitem{\unskip.}}
\providecommand*\mciteBstWouldAddEndPunctfalse
  {\let\EndOfBibitem\relax}
\providecommand*\mciteSetBstMidEndSepPunct[3]{}
\providecommand*\mciteSetBstSublistLabelBeginEnd[3]{}
\providecommand*\EndOfBibitem{}
\mciteSetBstSublistMode{f}
\mciteSetBstMaxWidthForm{subitem}{(\alph{mcitesubitemcount})}
\mciteSetBstSublistLabelBeginEnd
  {\mcitemaxwidthsubitemform\space}
  {\relax}
  {\relax}

\bibitem[Shibasaki and Yoshikawa(2002)Shibasaki, and Yoshikawa]{shibasaki2002lanthanide}
Shibasaki,~M.; Yoshikawa,~N. Lanthanide complexes in multifunctional asymmetric catalysis. \emph{Chemical reviews} \textbf{2002}, \emph{102}, 2187--2210\relax
\mciteBstWouldAddEndPuncttrue
\mciteSetBstMidEndSepPunct{\mcitedefaultmidpunct}
{\mcitedefaultendpunct}{\mcitedefaultseppunct}\relax
\EndOfBibitem
\bibitem[Woodruff \latin{et~al.}(2013)Woodruff, Winpenny, and Layfield]{woodruff2013lanthanide}
Woodruff,~D.~N.; Winpenny,~R.~E.; Layfield,~R.~A. Lanthanide single-molecule magnets. \emph{Chemical reviews} \textbf{2013}, \emph{113}, 5110--5148\relax
\mciteBstWouldAddEndPuncttrue
\mciteSetBstMidEndSepPunct{\mcitedefaultmidpunct}
{\mcitedefaultendpunct}{\mcitedefaultseppunct}\relax
\EndOfBibitem
\bibitem[Jin \latin{et~al.}(2026)Jin, Luo, and Zheng]{jin2026regulating}
Jin,~P.-B.; Luo,~Q.-C.; Zheng,~Y.-Z. Regulating Lanthanide Single-Molecule Magnets with Coordination Geometry and Organometallic Chemistry. \emph{Accounts of Chemical Research} \textbf{2026}, 2308548--28009\relax
\mciteBstWouldAddEndPuncttrue
\mciteSetBstMidEndSepPunct{\mcitedefaultmidpunct}
{\mcitedefaultendpunct}{\mcitedefaultseppunct}\relax
\EndOfBibitem
\bibitem[Li \latin{et~al.}(2025)Li, Zhao, Yu, and Zhan]{li2025recent}
Li,~J.; Zhao,~Y.; Yu,~D.; Zhan,~C. Recent Advances in d-f Transition Lanthanide Complexes for Organic Light-Emitting Diodes: Insights Into Structure--Luminescence Relationships. \emph{Laser \& Photonics Reviews} \textbf{2025}, \emph{19}, 2402198\relax
\mciteBstWouldAddEndPuncttrue
\mciteSetBstMidEndSepPunct{\mcitedefaultmidpunct}
{\mcitedefaultendpunct}{\mcitedefaultseppunct}\relax
\EndOfBibitem
\bibitem[Huizenga \latin{et~al.}(2021)Huizenga, Hratchian, and Jarrold]{huizenga2021lanthanide}
Huizenga,~C.; Hratchian,~H.~P.; Jarrold,~C.~C. Lanthanide oxides: From diatomics to high-spin, strongly correlated homo-and heterometallic clusters. \emph{The Journal of Physical Chemistry A} \textbf{2021}, \emph{125}, 6315--6331\relax
\mciteBstWouldAddEndPuncttrue
\mciteSetBstMidEndSepPunct{\mcitedefaultmidpunct}
{\mcitedefaultendpunct}{\mcitedefaultseppunct}\relax
\EndOfBibitem
\bibitem[Marin \latin{et~al.}(2021)Marin, Brunet, and Murugesu]{marin2021shining}
Marin,~R.; Brunet,~G.; Murugesu,~M. Shining new light on multifunctional lanthanide single-molecule magnets. \emph{Angewandte Chemie International Edition} \textbf{2021}, \emph{60}, 1728--1746\relax
\mciteBstWouldAddEndPuncttrue
\mciteSetBstMidEndSepPunct{\mcitedefaultmidpunct}
{\mcitedefaultendpunct}{\mcitedefaultseppunct}\relax
\EndOfBibitem
\bibitem[Nain and Ali(2024)Nain, and Ali]{nain2024modulation}
Nain,~S.; Ali,~M.~E. Modulation of the magnetic anisotropy via the ligand field in sandwiched erbium complexes. \emph{Inorganic Chemistry} \textbf{2024}, \emph{64}, 275--285\relax
\mciteBstWouldAddEndPuncttrue
\mciteSetBstMidEndSepPunct{\mcitedefaultmidpunct}
{\mcitedefaultendpunct}{\mcitedefaultseppunct}\relax
\EndOfBibitem
\bibitem[La~Macchia \latin{et~al.}(2008)La~Macchia, Infante, Raab, Gibson, and Gagliardi]{la2008theoretical}
La~Macchia,~G.; Infante,~I.; Raab,~J.; Gibson,~J.~K.; Gagliardi,~L. A theoretical study of the ground state and lowest excited states of PuO 0/+/+ 2 and PuO 2 0/+/+ 2. \emph{Physical Chemistry Chemical Physics} \textbf{2008}, \emph{10}, 7278--7283\relax
\mciteBstWouldAddEndPuncttrue
\mciteSetBstMidEndSepPunct{\mcitedefaultmidpunct}
{\mcitedefaultendpunct}{\mcitedefaultseppunct}\relax
\EndOfBibitem
\bibitem[Weichman \latin{et~al.}(2017)Weichman, Vlaisavljevich, DeVine, Shuman, Ard, Shiozaki, Neumark, and Viggiano]{weichman2017electronic}
Weichman,~M.~L.; Vlaisavljevich,~B.; DeVine,~J.~A.; Shuman,~N.~S.; Ard,~S.~G.; Shiozaki,~T.; Neumark,~D.~M.; Viggiano,~A.~A. Electronic structure of SmO and SmO- via slow photoelectron velocity-map imaging spectroscopy and spin-orbit CASPT2 calculations. \emph{The Journal of chemical physics} \textbf{2017}, \emph{147}\relax
\mciteBstWouldAddEndPuncttrue
\mciteSetBstMidEndSepPunct{\mcitedefaultmidpunct}
{\mcitedefaultendpunct}{\mcitedefaultseppunct}\relax
\EndOfBibitem
\bibitem[Coreno \latin{et~al.}(2006)Coreno, de~Simone, Green, Kaltsoyannis, Narband, and Sella]{coreno2006photoelectron}
Coreno,~M.; de~Simone,~M.; Green,~J.~C.; Kaltsoyannis,~N.; Narband,~N.; Sella,~A. Photoelectron spectroscopy of Ce ($\eta$-C5H5) 3--Accessing two ion states on 4f ionization. \emph{Chemical physics letters} \textbf{2006}, \emph{432}, 17--21\relax
\mciteBstWouldAddEndPuncttrue
\mciteSetBstMidEndSepPunct{\mcitedefaultmidpunct}
{\mcitedefaultendpunct}{\mcitedefaultseppunct}\relax
\EndOfBibitem
\bibitem[Babin \latin{et~al.}(2021)Babin, DeWitt, DeVine, McDonald, Ard, Shuman, Viggiano, Cheng, and Neumark]{babin2021electronic}
Babin,~M.~C.; DeWitt,~M.; DeVine,~J.~A.; McDonald,~D.~C.; Ard,~S.~G.; Shuman,~N.~S.; Viggiano,~A.~A.; Cheng,~L.; Neumark,~D.~M. Electronic structure of NdO via slow photoelectron velocity-map imaging spectroscopy of NdO---. \emph{The Journal of Chemical Physics} \textbf{2021}, \emph{155}\relax
\mciteBstWouldAddEndPuncttrue
\mciteSetBstMidEndSepPunct{\mcitedefaultmidpunct}
{\mcitedefaultendpunct}{\mcitedefaultseppunct}\relax
\EndOfBibitem
\bibitem[Linton \latin{et~al.}(1983)Linton, Dulick, Field, Carette, Leyland, and Barrow]{linton1983electronic}
Linton,~C.; Dulick,~M.; Field,~R.; Carette,~P.; Leyland,~P.; Barrow,~R. Electronic states of the CeO molecule: Absorption, emission, and laser spectroscopy. \emph{Journal of Molecular Spectroscopy} \textbf{1983}, \emph{102}, 441--497\relax
\mciteBstWouldAddEndPuncttrue
\mciteSetBstMidEndSepPunct{\mcitedefaultmidpunct}
{\mcitedefaultendpunct}{\mcitedefaultseppunct}\relax
\EndOfBibitem
\bibitem[Wyckoff and Wehinger(1977)Wyckoff, and Wehinger]{wyckoff1977ceo}
Wyckoff,~S.; Wehinger,~P. CeO-A new s-process molecule in S stars. \emph{Astrophysical Journal, Part 2-Letters to the Editor, vol. 212, Mar. 15, 1977, p. L139-L141. Research supported by the Science Research Council.} \textbf{1977}, \emph{212}, L139--L141\relax
\mciteBstWouldAddEndPuncttrue
\mciteSetBstMidEndSepPunct{\mcitedefaultmidpunct}
{\mcitedefaultendpunct}{\mcitedefaultseppunct}\relax
\EndOfBibitem
\bibitem[Moriyama \latin{et~al.}(2013)Moriyama, Tatewaki, and Yamamoto]{moriyama2013electronic}
Moriyama,~H.; Tatewaki,~H.; Yamamoto,~S. Electronic structure of CeO studied by a four-component relativistic configuration interaction method. \emph{The Journal of Chemical Physics} \textbf{2013}, \emph{138}\relax
\mciteBstWouldAddEndPuncttrue
\mciteSetBstMidEndSepPunct{\mcitedefaultmidpunct}
{\mcitedefaultendpunct}{\mcitedefaultseppunct}\relax
\EndOfBibitem
\bibitem[Fischer and Pratt(2022)Fischer, and Pratt]{fischer2022photoelectron}
Fischer,~I.; Pratt,~S.~T. Photoelectron spectroscopy in molecular physical chemistry. \emph{Physical Chemistry Chemical Physics} \textbf{2022}, \emph{24}, 1944--1959\relax
\mciteBstWouldAddEndPuncttrue
\mciteSetBstMidEndSepPunct{\mcitedefaultmidpunct}
{\mcitedefaultendpunct}{\mcitedefaultseppunct}\relax
\EndOfBibitem
\bibitem[Simons(2020)]{simons2020ejecting}
Simons,~J. Ejecting electrons from molecular anions via shine, shake/rattle, and roll. \emph{The Journal of Physical Chemistry A} \textbf{2020}, \emph{124}, 8778--8797\relax
\mciteBstWouldAddEndPuncttrue
\mciteSetBstMidEndSepPunct{\mcitedefaultmidpunct}
{\mcitedefaultendpunct}{\mcitedefaultseppunct}\relax
\EndOfBibitem
\bibitem[Mosaferi \latin{et~al.}(2022)Mosaferi, Selles, Miteva, Fert{\'e}, and Carniato]{mosaferi2022interpretation}
Mosaferi,~M.; Selles,~P.; Miteva,~T.; Fert{\'e},~A.; Carniato,~S. Interpretation of shakeup mechanisms in copper L-shell photoelectron spectra. \emph{The Journal of Physical Chemistry A} \textbf{2022}, \emph{126}, 4902--4914\relax
\mciteBstWouldAddEndPuncttrue
\mciteSetBstMidEndSepPunct{\mcitedefaultmidpunct}
{\mcitedefaultendpunct}{\mcitedefaultseppunct}\relax
\EndOfBibitem
\bibitem[Burroughs \latin{et~al.}(1976)Burroughs, Hamnett, Orchard, and Thornton]{burroughs1976satellite}
Burroughs,~P.; Hamnett,~A.; Orchard,~A.~F.; Thornton,~G. Satellite structure in the X-ray photoelectron spectra of some binary and mixed oxides of lanthanum and cerium. \emph{Journal of the Chemical Society, Dalton Transactions} \textbf{1976}, 1686--1698\relax
\mciteBstWouldAddEndPuncttrue
\mciteSetBstMidEndSepPunct{\mcitedefaultmidpunct}
{\mcitedefaultendpunct}{\mcitedefaultseppunct}\relax
\EndOfBibitem
\bibitem[Mason \latin{et~al.}(2019)Mason, Harb, Topolski, Hratchian, and Jarrold]{mason2019exceptionally}
Mason,~J.~L.; Harb,~H.; Topolski,~J.~E.; Hratchian,~H.~P.; Jarrold,~C.~C. Exceptionally complex electronic structures of lanthanide oxides and small molecules. \emph{Accounts of Chemical Research} \textbf{2019}, \emph{52}, 3265--3273\relax
\mciteBstWouldAddEndPuncttrue
\mciteSetBstMidEndSepPunct{\mcitedefaultmidpunct}
{\mcitedefaultendpunct}{\mcitedefaultseppunct}\relax
\EndOfBibitem
\bibitem[Teterin \latin{et~al.}(2004)Teterin, Teterin, Lebedev, and Ivanov]{teterin2004secondary}
Teterin,~Y.~A.; Teterin,~A.~Y.; Lebedev,~A.; Ivanov,~K. Secondary electronic processes and the structure of X-ray photoelectron spectra of lanthanides in oxygen-containing compounds. \emph{Journal of electron spectroscopy and related phenomena} \textbf{2004}, \emph{137}, 607--612\relax
\mciteBstWouldAddEndPuncttrue
\mciteSetBstMidEndSepPunct{\mcitedefaultmidpunct}
{\mcitedefaultendpunct}{\mcitedefaultseppunct}\relax
\EndOfBibitem
\bibitem[Zhou \latin{et~al.}(2022)Zhou, Hermes, Wu, Bao, Pandharkar, King, Zhang, Scott, Lykhin, Gagliardi, \latin{et~al.} others]{zhou2022electronic}
Zhou,~C.; Hermes,~M.~R.; Wu,~D.; Bao,~J.~J.; Pandharkar,~R.; King,~D.~S.; Zhang,~D.; Scott,~T.~R.; Lykhin,~A.~O.; Gagliardi,~L.; others Electronic structure of strongly correlated systems: recent developments in multiconfiguration pair-density functional theory and multiconfiguration nonclassical-energy functional theory. \emph{Chemical Science} \textbf{2022}, \emph{13}, 7685--7706\relax
\mciteBstWouldAddEndPuncttrue
\mciteSetBstMidEndSepPunct{\mcitedefaultmidpunct}
{\mcitedefaultendpunct}{\mcitedefaultseppunct}\relax
\EndOfBibitem
\bibitem[Nain \latin{et~al.}(2024)Nain, Mukhopadhyaya, and Ali]{nain2024unravelling}
Nain,~S.; Mukhopadhyaya,~A.; Ali,~M.~E. Unravelling the Highest Magnetic Anisotropy Among all the nd-Shells in [WCp 2] 0 Metallocene. \emph{Inorganic Chemistry} \textbf{2024}, \emph{63}, 7401--7411\relax
\mciteBstWouldAddEndPuncttrue
\mciteSetBstMidEndSepPunct{\mcitedefaultmidpunct}
{\mcitedefaultendpunct}{\mcitedefaultseppunct}\relax
\EndOfBibitem
\bibitem[Cao \latin{et~al.}(2021)Cao, Zhang, Wu, and Yang]{cao2021threshold}
Cao,~W.; Zhang,~Y.; Wu,~L.; Yang,~D.-S. Threshold ionization spectroscopy and theoretical calculations of LnO (Ln= La and Ce). \emph{The Journal of Physical Chemistry A} \textbf{2021}, \emph{125}, 1941--1948\relax
\mciteBstWouldAddEndPuncttrue
\mciteSetBstMidEndSepPunct{\mcitedefaultmidpunct}
{\mcitedefaultendpunct}{\mcitedefaultseppunct}\relax
\EndOfBibitem
\bibitem[Ray \latin{et~al.}(2015)Ray, Felton, Kafader, Topolski, and Jarrold]{ray2015photoelectron}
Ray,~M.; Felton,~J.~A.; Kafader,~J.~O.; Topolski,~J.~E.; Jarrold,~C.~C. Photoelectron spectra of CeO- and Ce (OH) 2-. \emph{The Journal of Chemical Physics} \textbf{2015}, \emph{142}\relax
\mciteBstWouldAddEndPuncttrue
\mciteSetBstMidEndSepPunct{\mcitedefaultmidpunct}
{\mcitedefaultendpunct}{\mcitedefaultseppunct}\relax
\EndOfBibitem
\bibitem[Moravec and Jarrold(1998)Moravec, and Jarrold]{moravec1998study}
Moravec,~V.~D.; Jarrold,~C.~C. Study of the low-lying states of NiO- and NiO using anion photoelectron spectroscopy. \emph{The Journal of chemical physics} \textbf{1998}, \emph{108}, 1804--1810\relax
\mciteBstWouldAddEndPuncttrue
\mciteSetBstMidEndSepPunct{\mcitedefaultmidpunct}
{\mcitedefaultendpunct}{\mcitedefaultseppunct}\relax
\EndOfBibitem
\bibitem[Waller \latin{et~al.}(2012)Waller, Mann, Rothgeb, and Jarrold]{waller2012study}
Waller,~S.~E.; Mann,~J.~E.; Rothgeb,~D.~W.; Jarrold,~C.~C. Study of MoNbO y (y= 2--5) Anion and Neutral Clusters using Photoelectron Spectroscopy and Density Functional Theory Calculations: Impact of Spin Contamination on Single Point Calculations. \emph{The Journal of Physical Chemistry A} \textbf{2012}, \emph{116}, 9639--9652\relax
\mciteBstWouldAddEndPuncttrue
\mciteSetBstMidEndSepPunct{\mcitedefaultmidpunct}
{\mcitedefaultendpunct}{\mcitedefaultseppunct}\relax
\EndOfBibitem
\bibitem[Felton \latin{et~al.}(2014)Felton, Ray, and Jarrold]{felton2014measurement}
Felton,~J.; Ray,~M.; Jarrold,~C.~C. Measurement of the electron affinity of atomic Ce. \emph{Physical Review A} \textbf{2014}, \emph{89}, 033407\relax
\mciteBstWouldAddEndPuncttrue
\mciteSetBstMidEndSepPunct{\mcitedefaultmidpunct}
{\mcitedefaultendpunct}{\mcitedefaultseppunct}\relax
\EndOfBibitem
\bibitem[Malmqvist and Roos(1989)Malmqvist, and Roos]{malmqvist1989casscf}
Malmqvist,~P.-{\AA}.; Roos,~B.~O. The CASSCF state interaction method. \emph{Chemical physics letters} \textbf{1989}, \emph{155}, 189--194\relax
\mciteBstWouldAddEndPuncttrue
\mciteSetBstMidEndSepPunct{\mcitedefaultmidpunct}
{\mcitedefaultendpunct}{\mcitedefaultseppunct}\relax
\EndOfBibitem
\bibitem[Angeli \latin{et~al.}(2001)Angeli, Cimiraglia, Evangelisti, Leininger, and Malrieu]{angeli2001introduction}
Angeli,~C.; Cimiraglia,~R.; Evangelisti,~S.; Leininger,~T.; Malrieu,~J.-P. Introduction of n-electron valence states for multireference perturbation theory. \emph{The Journal of Chemical Physics} \textbf{2001}, \emph{114}, 10252--10264\relax
\mciteBstWouldAddEndPuncttrue
\mciteSetBstMidEndSepPunct{\mcitedefaultmidpunct}
{\mcitedefaultendpunct}{\mcitedefaultseppunct}\relax
\EndOfBibitem
\bibitem[He{\ss} \latin{et~al.}(1996)He{\ss}, Marian, Wahlgren, and Gropen]{hess1996mean}
He{\ss},~B.~A.; Marian,~C.~M.; Wahlgren,~U.; Gropen,~O. A mean-field spin-orbit method applicable to correlated wavefunctions. \emph{Chemical Physics Letters} \textbf{1996}, \emph{251}, 365--371\relax
\mciteBstWouldAddEndPuncttrue
\mciteSetBstMidEndSepPunct{\mcitedefaultmidpunct}
{\mcitedefaultendpunct}{\mcitedefaultseppunct}\relax
\EndOfBibitem
\bibitem[Angeli \latin{et~al.}(2004)Angeli, Borini, Cestari, and Cimiraglia]{angeli2004quasidegenerate}
Angeli,~C.; Borini,~S.; Cestari,~M.; Cimiraglia,~R. A quasidegenerate formulation of the second order n-electron valence state perturbation theory approach. \emph{The Journal of chemical physics} \textbf{2004}, \emph{121}, 4043--4049\relax
\mciteBstWouldAddEndPuncttrue
\mciteSetBstMidEndSepPunct{\mcitedefaultmidpunct}
{\mcitedefaultendpunct}{\mcitedefaultseppunct}\relax
\EndOfBibitem
\bibitem[Weigend and Ahlrichs(2005)Weigend, and Ahlrichs]{weigend2005balanced}
Weigend,~F.; Ahlrichs,~R. Balanced basis sets of split valence, triple zeta valence and quadruple zeta valence quality for H to Rn: Design and assessment of accuracy. \emph{Physical Chemistry Chemical Physics} \textbf{2005}, \emph{7}, 3297--3305\relax
\mciteBstWouldAddEndPuncttrue
\mciteSetBstMidEndSepPunct{\mcitedefaultmidpunct}
{\mcitedefaultendpunct}{\mcitedefaultseppunct}\relax
\EndOfBibitem
\bibitem[Nakajima and Hirao(2012)Nakajima, and Hirao]{nakajima2012douglas}
Nakajima,~T.; Hirao,~K. The Douglas--Kroll--Hess Approach. \emph{Chemical reviews} \textbf{2012}, \emph{112}, 385--402\relax
\mciteBstWouldAddEndPuncttrue
\mciteSetBstMidEndSepPunct{\mcitedefaultmidpunct}
{\mcitedefaultendpunct}{\mcitedefaultseppunct}\relax
\EndOfBibitem
\bibitem[Neese(2025)]{neese2025software}
Neese,~F. Software update: the ORCA program system—version 6.0. \emph{Wiley Interdisciplinary Reviews: Computational Molecular Science} \textbf{2025}, \emph{15}, e70019\relax
\mciteBstWouldAddEndPuncttrue
\mciteSetBstMidEndSepPunct{\mcitedefaultmidpunct}
{\mcitedefaultendpunct}{\mcitedefaultseppunct}\relax
\EndOfBibitem
\bibitem[Reid(2003)]{reid2003photoelectron}
Reid,~K.~L. Photoelectron angular distributions. \emph{Annual review of physical chemistry} \textbf{2003}, \emph{54}, 397--424\relax
\mciteBstWouldAddEndPuncttrue
\mciteSetBstMidEndSepPunct{\mcitedefaultmidpunct}
{\mcitedefaultendpunct}{\mcitedefaultseppunct}\relax
\EndOfBibitem
\bibitem[Sanov(2014)]{sanov2014laboratory}
Sanov,~A. Laboratory-frame photoelectron angular distributions in anion photodetachment: insight into electronic structure and intermolecular interactions. \emph{Annual review of physical chemistry} \textbf{2014}, \emph{65}, 341--363\relax
\mciteBstWouldAddEndPuncttrue
\mciteSetBstMidEndSepPunct{\mcitedefaultmidpunct}
{\mcitedefaultendpunct}{\mcitedefaultseppunct}\relax
\EndOfBibitem
\bibitem[Anstoter and Verlet(2021)Anstoter, and Verlet]{anstoter2021modeling}
Anstoter,~C.~S.; Verlet,~J.~R. Modeling the photoelectron angular distributions of molecular anions: roles of the basis set, orbital choice, and geometry. \emph{The Journal of Physical Chemistry A} \textbf{2021}, \emph{125}, 4888--4895\relax
\mciteBstWouldAddEndPuncttrue
\mciteSetBstMidEndSepPunct{\mcitedefaultmidpunct}
{\mcitedefaultendpunct}{\mcitedefaultseppunct}\relax
\EndOfBibitem
\bibitem[Tufekci \latin{et~al.}(2024)Tufekci, Foreman, Romeu, Dixon, Peterson, Cheng, and Bowen]{tufekci2024anion}
Tufekci,~B.~A.; Foreman,~K.; Romeu,~J.~G.; Dixon,~D.~A.; Peterson,~K.~A.; Cheng,~L.; Bowen,~K.~H. Anion Photoelectron Spectroscopy and Ab Initio Studies of the UF--Anion. \emph{The Journal of Physical Chemistry Letters} \textbf{2024}, \emph{15}, 11932--11938\relax
\mciteBstWouldAddEndPuncttrue
\mciteSetBstMidEndSepPunct{\mcitedefaultmidpunct}
{\mcitedefaultendpunct}{\mcitedefaultseppunct}\relax
\EndOfBibitem
\bibitem[Xie and Zare(1990)Xie, and Zare]{Xie.1990}
Xie,~J.; Zare,~R.~N. {Selection rules for the photoionization of diatomic molecules}. \emph{The Journal of Chemical Physics} \textbf{1990}, \emph{93}, 3033--3038, DOI: \doi{10.1063/1.458837}\relax
\mciteBstWouldAddEndPuncttrue
\mciteSetBstMidEndSepPunct{\mcitedefaultmidpunct}
{\mcitedefaultendpunct}{\mcitedefaultseppunct}\relax
\EndOfBibitem
\bibitem[Moravec \latin{et~al.}(1999)Moravec, Klopcic, and Jarrold]{Moravec.1999}
Moravec,~V.~D.; Klopcic,~S.~A.; Jarrold,~C.~C. {Anion photoelectron spectroscopy of small tin clusters}. \emph{The Journal of Chemical Physics} \textbf{1999}, \emph{110}, 5079--5088, DOI: \doi{10.1063/1.478405}\relax
\mciteBstWouldAddEndPuncttrue
\mciteSetBstMidEndSepPunct{\mcitedefaultmidpunct}
{\mcitedefaultendpunct}{\mcitedefaultseppunct}\relax
\EndOfBibitem
\bibitem[Manson and Starace(1982)Manson, and Starace]{RevModPhys.54.389}
Manson,~S.~T.; Starace,~A.~F. Photoelectron angular distributions: energy dependence for $s$ subshells. \emph{Rev. Mod. Phys.} \textbf{1982}, \emph{54}, 389--405, DOI: \doi{10.1103/RevModPhys.54.389}\relax
\mciteBstWouldAddEndPuncttrue
\mciteSetBstMidEndSepPunct{\mcitedefaultmidpunct}
{\mcitedefaultendpunct}{\mcitedefaultseppunct}\relax
\EndOfBibitem
\bibitem[Amusia(1975)]{Amusia1975}
Amusia,~M.~Y. Many-Electron Description of Atomic Photoeffect. \emph{Phys. Rep.} \textbf{1975}, \emph{22}, 263--315\relax
\mciteBstWouldAddEndPuncttrue
\mciteSetBstMidEndSepPunct{\mcitedefaultmidpunct}
{\mcitedefaultendpunct}{\mcitedefaultseppunct}\relax
\EndOfBibitem
\bibitem[Brundle \latin{et~al.}(1979)Brundle, Baker, and Thomas]{brundle1979electron}
Brundle,~C.~R.; Baker,~A.~D.; Thomas,~T.~D. Electron Spectroscopy: Theory, Techniques, and Applications, Volume 2. 1979\relax
\mciteBstWouldAddEndPuncttrue
\mciteSetBstMidEndSepPunct{\mcitedefaultmidpunct}
{\mcitedefaultendpunct}{\mcitedefaultseppunct}\relax
\EndOfBibitem
\bibitem[Starace(1982)]{starace1982theory}
Starace,~A.~F. Theory of atomic photoionization. \emph{Handbuch der Physik} \textbf{1982}, \emph{6}, 1--121\relax
\mciteBstWouldAddEndPuncttrue
\mciteSetBstMidEndSepPunct{\mcitedefaultmidpunct}
{\mcitedefaultendpunct}{\mcitedefaultseppunct}\relax
\EndOfBibitem
\bibitem[Thomas(1984)]{thomas1984transition}
Thomas,~T.~D. Transition from adiabatic to sudden excitation of core electrons. \emph{Physical review letters} \textbf{1984}, \emph{52}, 417\relax
\mciteBstWouldAddEndPuncttrue
\mciteSetBstMidEndSepPunct{\mcitedefaultmidpunct}
{\mcitedefaultendpunct}{\mcitedefaultseppunct}\relax
\EndOfBibitem
\bibitem[Ungier and Thomas(1984)Ungier, and Thomas]{ungier1984resonance}
Ungier,~L.; Thomas,~T. Resonance-enhanced shakeup in near-threshold core excitation of CO and N 2. \emph{Physical review letters} \textbf{1984}, \emph{53}, 435\relax
\mciteBstWouldAddEndPuncttrue
\mciteSetBstMidEndSepPunct{\mcitedefaultmidpunct}
{\mcitedefaultendpunct}{\mcitedefaultseppunct}\relax
\EndOfBibitem
\bibitem[Schirmer \latin{et~al.}(1991)Schirmer, Braunstein, and McKoy]{schirmer1991satellite}
Schirmer,~J.; Braunstein,~M.; McKoy,~V. Satellite intensities in the K-shell photoionization of CO. \emph{Physical Review A} \textbf{1991}, \emph{44}, 5762\relax
\mciteBstWouldAddEndPuncttrue
\mciteSetBstMidEndSepPunct{\mcitedefaultmidpunct}
{\mcitedefaultendpunct}{\mcitedefaultseppunct}\relax
\EndOfBibitem
\bibitem[Fujiwara \latin{et~al.}(2005)Fujiwara, Pr{\"u}mper, De~Fanis, Tamenori, Tanaka, Kitajima, Tanaka, Oura, and Ueda]{fujiwara2005excitation}
Fujiwara,~K.; Pr{\"u}mper,~G.; De~Fanis,~A.; Tamenori,~Y.; Tanaka,~T.; Kitajima,~M.; Tanaka,~H.; Oura,~M.; Ueda,~K. The excitation mechanism of satellite bands in F 1s photoemission of SiF4. \emph{Chemical physics letters} \textbf{2005}, \emph{402}, 17--20\relax
\mciteBstWouldAddEndPuncttrue
\mciteSetBstMidEndSepPunct{\mcitedefaultmidpunct}
{\mcitedefaultendpunct}{\mcitedefaultseppunct}\relax
\EndOfBibitem
\bibitem[Bilodeau \latin{et~al.}(2012)Bilodeau, Gibson, Walter, Aguilar, and Berrah]{Bilodeau.2012}
Bilodeau,~R.; Gibson,~N.; Walter,~C.; Aguilar,~A.; Berrah,~N. {Inner-shell photodetachment: Shape and Feshbach resonances of anions}. \emph{Journal of Electron Spectroscopy and Related Phenomena} \textbf{2012}, \emph{185}, 219--225, DOI: \doi{10.1016/j.elspec.2012.06.015}\relax
\mciteBstWouldAddEndPuncttrue
\mciteSetBstMidEndSepPunct{\mcitedefaultmidpunct}
{\mcitedefaultendpunct}{\mcitedefaultseppunct}\relax
\EndOfBibitem
\bibitem[Mason \latin{et~al.}(2021)Mason, Harb, Taka, McMahon, Huizenga, Corzo, Hratchian, and Jarrold]{mason2021photoelectron}
Mason,~J.~L.; Harb,~H.; Taka,~A.~A.; McMahon,~A.~J.; Huizenga,~C.~D.; Corzo,~H.; Hratchian,~H.~P.; Jarrold,~C.~C. Photoelectron Spectra of Gd2O2--and Nonmonotonic Photon-Energy-Dependent Variations in Populations of Close-Lying Neutral States. \emph{The Journal of Physical Chemistry A} \textbf{2021}, \emph{125}, 857--866\relax
\mciteBstWouldAddEndPuncttrue
\mciteSetBstMidEndSepPunct{\mcitedefaultmidpunct}
{\mcitedefaultendpunct}{\mcitedefaultseppunct}\relax
\EndOfBibitem
\bibitem[Mason \latin{et~al.}(2021)Mason, Harb, Taka, Huizenga, Corzo, Hratchian, and Jarrold]{Mason2021}
Mason,~J.~L.; Harb,~H.; Taka,~A.~A.; Huizenga,~C.~D.; Corzo,~H.~H.; Hratchian,~H.~P.; Jarrold,~C.~C. New Photoelectron–Valence Electron Interactions Evident in the Photoelectron Spectrum of Gd2O–. \emph{The Journal of Physical Chemistry A} \textbf{2021}, \emph{125}, 9892--9903, DOI: \doi{10.1021/acs.jpca.1c07818}\relax
\mciteBstWouldAddEndPuncttrue
\mciteSetBstMidEndSepPunct{\mcitedefaultmidpunct}
{\mcitedefaultendpunct}{\mcitedefaultseppunct}\relax
\EndOfBibitem
\bibitem[Huizenga \latin{et~al.}(2025)Huizenga, Vaish, Dwyer, Taka, Bovill, Thompson, Hratchian, and Jarrold]{Huizenga.202555o}
Huizenga,~C.~D.; Vaish,~S.; Dwyer,~C.; Taka,~A.~A.; Bovill,~A.~J.; Thompson,~L.~M.; Hratchian,~H.~P.; Jarrold,~C.~C. {Exploring Anomalous Photoelectron Angular Distributions in the Photoelectron Spectra of Gd3O3 –: Study of Gd3O2 – and Gd3O3 – Using Photoelectron Spectroscopy and Density Functional Theory Calculations}. \emph{The Journal of Physical Chemistry A} \textbf{2025}, DOI: \doi{10.1021/acs.jpca.5c06858}\relax
\mciteBstWouldAddEndPuncttrue
\mciteSetBstMidEndSepPunct{\mcitedefaultmidpunct}
{\mcitedefaultendpunct}{\mcitedefaultseppunct}\relax
\EndOfBibitem
\bibitem[Huizenga \latin{et~al.}(2025)Huizenga, Vaish, Thompson, and Jarrold]{Huizenga.2025}
Huizenga,~C.~D.; Vaish,~S.; Thompson,~L.~M.; Jarrold,~C.~C. {Electronic structures and spin frustration in Ln3O (Ln = Ce, Sm, Gd) neutrals and anions determined by anion photoelectron spectroscopy}. \emph{The Journal of Chemical Physics} \textbf{2025}, \emph{162}, 054301, DOI: \doi{10.1063/5.0249692}\relax
\mciteBstWouldAddEndPuncttrue
\mciteSetBstMidEndSepPunct{\mcitedefaultmidpunct}
{\mcitedefaultendpunct}{\mcitedefaultseppunct}\relax
\EndOfBibitem
\bibitem[Kinyua \latin{et~al.}(2025)Kinyua, Hratchian, Jarrold, and Thompson]{Kinyua.2025}
Kinyua,~A.~M.; Hratchian,~H.~P.; Jarrold,~C.~C.; Thompson,~L.~M. {Photoelectron–remnant interaction effect on remnant wavefunction in low-kinetic energy electron detachment events}. \emph{The Journal of Chemical Physics} \textbf{2025}, \emph{162}, 064304, DOI: \doi{10.1063/5.0245067}\relax
\mciteBstWouldAddEndPuncttrue
\mciteSetBstMidEndSepPunct{\mcitedefaultmidpunct}
{\mcitedefaultendpunct}{\mcitedefaultseppunct}\relax
\EndOfBibitem
\bibitem[Cooper and Zare(1968)Cooper, and Zare]{cooper1968angular}
Cooper,~J.; Zare,~R. Angular distribution of photoelectrons. \emph{J. Chem. Phys.} \textbf{1968}, \emph{48}\relax
\mciteBstWouldAddEndPuncttrue
\mciteSetBstMidEndSepPunct{\mcitedefaultmidpunct}
{\mcitedefaultendpunct}{\mcitedefaultseppunct}\relax
\EndOfBibitem
\bibitem[Mishra \latin{et~al.}(2006)Mishra, Vallet, and Domcke]{mishra2006importance}
Mishra,~S.; Vallet,~V.; Domcke,~W. Importance of spin--orbit Coupling for the Assignment of the Photodetachment Spectra of AuX2-(X= Cl, Br, and I). \emph{ChemPhysChem} \textbf{2006}, \emph{7}, 723--727\relax
\mciteBstWouldAddEndPuncttrue
\mciteSetBstMidEndSepPunct{\mcitedefaultmidpunct}
{\mcitedefaultendpunct}{\mcitedefaultseppunct}\relax
\EndOfBibitem
\bibitem[Cantero-L{\'o}pez \latin{et~al.}(2021)Cantero-L{\'o}pez, Santoyo-Flores, Vega, Carre{\~n}o, Fuentes, Ramirez-Osorio, Ortiz, Illicachi, S{\'a}nchez, Olea, \latin{et~al.} others]{cantero2021theoretical}
Cantero-L{\'o}pez,~P.; Santoyo-Flores,~J.; Vega,~A.; Carre{\~n}o,~A.; Fuentes,~J.~A.; Ramirez-Osorio,~A.; Ortiz,~A.; Illicachi,~L.~A.; S{\'a}nchez,~J.; Olea,~A.~F.; others A theoretical chemistry-based strategy for the rational design of new luminescent lanthanide complexes: an approach from a multireference SOC-NEVPT2 method. \emph{Dalton Transactions} \textbf{2021}, \emph{50}, 13561--13571\relax
\mciteBstWouldAddEndPuncttrue
\mciteSetBstMidEndSepPunct{\mcitedefaultmidpunct}
{\mcitedefaultendpunct}{\mcitedefaultseppunct}\relax
\EndOfBibitem
\bibitem[Makhlouf \latin{et~al.}(2021)Makhlouf, Adem, Magnier, and Taher]{makhlouf2021theoretical}
Makhlouf,~S.; Adem,~Z.; Magnier,~S.; Taher,~F. Theoretical spin-orbit calculations of low-lying electronic states of cerium monoxide. \emph{Journal of Quantitative Spectroscopy and Radiative Transfer} \textbf{2021}, \emph{276}, 107894\relax
\mciteBstWouldAddEndPuncttrue
\mciteSetBstMidEndSepPunct{\mcitedefaultmidpunct}
{\mcitedefaultendpunct}{\mcitedefaultseppunct}\relax
\EndOfBibitem
\bibitem[Kaledin \latin{et~al.}(1993)Kaledin, McCord, and Heaven]{kaledin1993laser}
Kaledin,~L.~A.; McCord,~J.; Heaven,~M.~C. Laser spectroscopy of CeO: characterization and assignment of states in the 0-3 eV range. \emph{Journal of Molecular Spectroscopy} \textbf{1993}, \emph{158}, 40--61\relax
\mciteBstWouldAddEndPuncttrue
\mciteSetBstMidEndSepPunct{\mcitedefaultmidpunct}
{\mcitedefaultendpunct}{\mcitedefaultseppunct}\relax
\EndOfBibitem
\bibitem[Angeli \latin{et~al.}(2001)Angeli, Cimiraglia, and Malrieu]{angeli2001n}
Angeli,~C.; Cimiraglia,~R.; Malrieu,~J.-P. N-electron valence state perturbation theory: a fast implementation of the strongly contracted variant. \emph{Chemical physics letters} \textbf{2001}, \emph{350}, 297--305\relax
\mciteBstWouldAddEndPuncttrue
\mciteSetBstMidEndSepPunct{\mcitedefaultmidpunct}
{\mcitedefaultendpunct}{\mcitedefaultseppunct}\relax
\EndOfBibitem
\bibitem[Lischka \latin{et~al.}(2018)Lischka, Nachtigallova, Aquino, Szalay, Plasser, Machado, and Barbatti]{lischka2018multireference}
Lischka,~H.; Nachtigallova,~D.; Aquino,~A.~J.; Szalay,~P.~G.; Plasser,~F.; Machado,~F.~B.; Barbatti,~M. Multireference approaches for excited states of molecules. \emph{Chemical reviews} \textbf{2018}, \emph{118}, 7293--7361\relax
\mciteBstWouldAddEndPuncttrue
\mciteSetBstMidEndSepPunct{\mcitedefaultmidpunct}
{\mcitedefaultendpunct}{\mcitedefaultseppunct}\relax
\EndOfBibitem
\bibitem[Oana and Krylov(2009)Oana, and Krylov]{oana2009cross}
Oana,~C.~M.; Krylov,~A.~I. Cross sections and photoelectron angular distributions in photodetachment from negative ions using equation-of-motion coupled-cluster Dyson orbitals. \emph{The Journal of chemical physics} \textbf{2009}, \emph{131}\relax
\mciteBstWouldAddEndPuncttrue
\mciteSetBstMidEndSepPunct{\mcitedefaultmidpunct}
{\mcitedefaultendpunct}{\mcitedefaultseppunct}\relax
\EndOfBibitem
\bibitem[Tenorio \latin{et~al.}(2022)Tenorio, Ponzi, Coriani, and Decleva]{tenorio2022photoionization}
Tenorio,~B. N.~C.; Ponzi,~A.; Coriani,~S.; Decleva,~P. Photoionization observables from multi-reference Dyson orbitals coupled to B-spline DFT and TD-DFT continuum. \emph{Molecules} \textbf{2022}, \emph{27}, 1203\relax
\mciteBstWouldAddEndPuncttrue
\mciteSetBstMidEndSepPunct{\mcitedefaultmidpunct}
{\mcitedefaultendpunct}{\mcitedefaultseppunct}\relax
\EndOfBibitem
\bibitem[Xie and Zare(1990)Xie, and Zare]{xie1990selection}
Xie,~J.; Zare,~R.~N. Selection rules for the photoionization of diatomic molecules. \emph{The Journal of chemical physics} \textbf{1990}, \emph{93}, 3033--3038\relax
\mciteBstWouldAddEndPuncttrue
\mciteSetBstMidEndSepPunct{\mcitedefaultmidpunct}
{\mcitedefaultendpunct}{\mcitedefaultseppunct}\relax
\EndOfBibitem
\bibitem[Magoulas \latin{et~al.}(2015)Magoulas, Papakondylis, and Mavridis]{magoulas2015structural}
Magoulas,~I.; Papakondylis,~A.; Mavridis,~A. Structural parameters of the ground states of the quasi-stable diatomic anions CO-, BF-, and BCl- as obtained by conventional Ab Initio methods. \emph{International Journal of Quantum Chemistry} \textbf{2015}, \emph{115}, 771--778\relax
\mciteBstWouldAddEndPuncttrue
\mciteSetBstMidEndSepPunct{\mcitedefaultmidpunct}
{\mcitedefaultendpunct}{\mcitedefaultseppunct}\relax
\EndOfBibitem
\bibitem[Jagau \latin{et~al.}(2017)Jagau, Bravaya, and Krylov]{jagau2017extending}
Jagau,~T.-C.; Bravaya,~K.~B.; Krylov,~A.~I. Extending quantum chemistry of bound states to electronic resonances. \emph{Annual review of physical chemistry} \textbf{2017}, \emph{68}, 525--553\relax
\mciteBstWouldAddEndPuncttrue
\mciteSetBstMidEndSepPunct{\mcitedefaultmidpunct}
{\mcitedefaultendpunct}{\mcitedefaultseppunct}\relax
\EndOfBibitem
\end{mcitethebibliography}

\end{document}


\section{DFT Results} \label{DFT_SI}
The molecular geometries of CeO and CeO$^{-}$ have been optimized at the density functional theory (DFT) level using the B3LYP functional in conjunction with the ZORA relativistic Hamiltonian.\cite{van1999geometry} For oxygen, the Ahlrichs def2-TZVP basis set with polarization functions\cite{weigend2005balanced} is employed, while for cerium a segmented all-electron relativistic contracted (SARC) triple-zeta basis set has been used to account for scalar relativistic effects.\cite{pantazis2009all} The AUTOAUX keyword was employed to automatically generate the corresponding auxiliary basis sets.\cite{stoychev2017automatic} Grimme’s D3 dispersion correction with Becke–Johnson damping was included to account for long-range correlation effects and to ensure a balanced description of weak interactions during geometry optimization.\cite{grimme2011effect}

\begin{table}[!ht]
\centering
\caption{Single-point electronic energies (E$_h$) of different spin states of CeO and CeO$^{-}$ computed at the B3LYP-D3/ZORA/def2-TZVP level of theory.}
\begin{tabular}{lll}
\hline
Molecule & Spin state & Energy (E$_h$) \\
\hline
CeO     & Singlet    & -9097.60755 \\
        & Triplet    & -9097.64269 \\
        & Quintet    & -9097.47943 \\
\hline
CeO$^{-}$ & Doublet  & -9097.67393 \\
          & Quartet  & -9097.64328 \\
\hline
\end{tabular}
\end{table}

\begin{figure}[!ht]
\centering
\includegraphics[width=6cm]{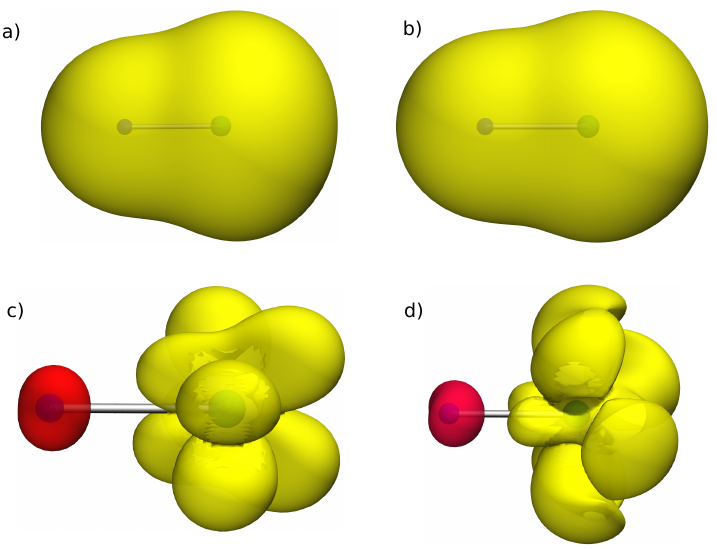}
 \caption{ Electron density distributions for (a) CeO$^{-}$, (b) CeO and spin density distributions for (c) CeO$^{-}$, (d) CeO, calculated at the B3LYP-D3/ZORA/def2-TZVP level. Isosurfaces are plotted at 0.003 e/Å$^{-3}$ (electron density) and 0.003 $\mu_{\mathrm{B}}$/Å$^{-3}$ (spin density). } 
   \label{fig:density_plot}
\end{figure}

The electronic structure of CeO is dominated by a dense manifold of Ce \(4f\) and \(5d\) states, leading to strong multiconfigurational character and significant spin-orbit coupling.\cite{makhlouf2021theoretical} Hybridization with O \(2p\) orbitals further shapes the bonding and electronic structure.\cite{maslakov2018electronic} As shown in Figure~\ref{fig:density_plot}, oxidation of \ce{CeO^-} produces only modest changes in the electron density, which remains largely localized on the Ce center and polarized along the Ce-O bond, consistent with predominantly ionic bonding. In contrast, the spin density is substantially redistributed. Neutral CeO exhibits spin density localized on Ce, while \ce{CeO^-} shows a more diffuse spin distribution, reflecting altered occupation within the Ce-centered \(4f/5d\) manifold. This behavior underscores the multiconfigurational nature of the electronic structure and the active role of Ce-centered orbitals in the photodetachment process.\cite{schafer2021cerium} The associated changes in frontier-orbital composition suggest subtle variations in the ligand field and magnetic anisotropy upon oxidation.
\begin{figure}[!ht]
\centering
\includegraphics[width=10cm]{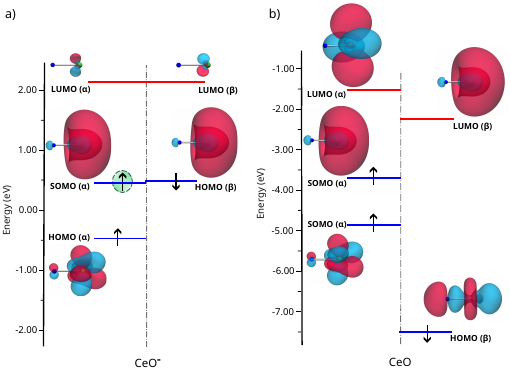}
 \caption{Representative spin-resolved MO energy levels and selected orbital isosurfaces of (a) CeO$^-$, (b) CeO, computed at the B3LYP-D3/ZORA-def2-TZVP level and plotted at an isovalue of 0.03. Occupied and unoccupied orbitals are shown in blue and red, respectively. For CeO$^-$ (doublet), the SOMO ($\alpha$) is Ce $6s$ while neutral CeO (triplet) has two SOMOs ($\alpha$) of Ce $4f$ and $6s$ character. The Ce $6s$ SOMO (detachment orbital) is highlighted in green.}
   \label{fig:dft_orbs}
\end{figure}
The frontier orbital ordering reflects the near-degeneracy of the Ce \(4f\), \(5d\), and \(6s\) valence shells. In \ce{CeO^-}, a singly occupied, largely nonbonding Ce \(4f\) orbital lies below a doubly occupied Ce \(6s\) orbital, while the low-lying virtual space is composed of closely spaced Ce \(5d_{\pi}\)/\(6p_{\pi}\) orbitals  (Figure \ref{fig:levels}). This non-Aufbau ordering arises from the small energy separations among the Ce valence orbitals, a hallmark of many transition-metal and lanthanide systems.\cite{westcott2000experimental,bunting2018linear,nain2024unravelling} In neutral CeO, the unpaired electrons occupy the Ce \(4f\) and \(6s\) orbitals. In both species, the Ce \(4f\) orbitals remain largely localized and nonbonding, whereas bonding is dominated by O \(2p\)--Ce \(5d\) interactions, yielding a predominantly metal-centered electronic structure. The dominant photodetachment channel originates from the Ce \(6s\)-based SOMO (Figure \ref{fig:dft_orbs}). Its diffuse, weakly bonding character minimizes geometric relaxation upon detachment, giving rise to compact Franck-Condon envelopes. 

\section{Active Space Benchmarking} \label{active space benchmarking}
The CAS(2,13) active space has been adopted as the optimal balance between accuracy and computational cost (Table~\ref{tab:active space benchmarking}). Although CAS(2,8) neglects important \(4f\)--\(5d\) interactions and compresses the excitation spectrum, inclusion of the Ce \(5d\) orbitals yields a more balanced description of the low-lying \(\Phi\), \(\Delta\), \(\Pi\), and \(\Sigma\) states. Further expansion to CAS(2,16) or CAS(8,16) produces only minor changes in excitation energies and state ordering, despite substantially increased computational expense. The low-lying electronic structure is therefore largely converged at the CAS(2,13) level, which is used throughout this work. 
\begin{table}[!ht]
\centering
\caption{Spin-free excitation energies (cm$^{-1}$) of the low-lying $\Phi$, $\Delta$, $\Pi$, and $\Sigma^+$ triplets of CeO relative to the ground state computed using CASSCF and NEVPT2 with different active spaces, averaging over seven triplet and seven singlet states.}
\label{tab:active space benchmarking}
\begin{tabular}{llllllll}
\toprule
{\bf{Method }}                        & $\bm{\Phi}$ & $\bm{\Delta}$ & $\bm{\Pi}$   & $\bm{\Sigma^{+}}$ \\
\hline
\multicolumn{5}{c}{{\bf{CAS(2,8)}} orb.= 4$f$+6$s$}  \\  
\hline
CASSCF                         & 0.00   & 668   & 1690 & 1713  \\
NEVPT2                         & 0.00    & 547   & 1031 & 530    \\
\hline
\multicolumn{5}{c}{{\bf{CAS(2,13)}} orb.= 4$f$+6$s$+5$d$} \\
\hline
CASSCF                         & 0.00    & 902   & 1947 & 1981   \\
NEVPT2                         & 0.00    & 670   & 1086 & 492    \\
\hline
\multicolumn{5}{c} {{\bf{CAS(2,16)}} orb.= 4$f$+6$s$+5$d$+6$p$(Ce)}           \\
\hline
CASSCF & 0.00    & 659   & 1680 & 1707   \\
NEVPT2                         & 0.00    & 542   & 1010 & 495    \\
\hline
\multicolumn{5}{c}{{\bf{CAS(8,16)}} orb.= 4$f$+6$s$+5$d$+2$p$(O)}  \\
\hline
CASSCF                         & 0.00    & 559   & 1613 & 1472   \\
NEVPT2                         & 0.00    & 659   & 1279 & 743  \\
\bottomrule
\end{tabular}
\end{table}
Both DFT and multireference calculations consistently predict a doublet ground state for \ce{CeO^-} and a triplet ground state for CeO. At the DFT level, including zero-point vibrational corrections, the adiabatic and vertical detachment energies are calculated to be 0.86 and 0.87 eV, respectively (Table~\ref{tab:ADE_VDE}). These values are in excellent agreement with the theoretical results of Ray et al., who reported ADE and VDE values of 0.87 and 0.89 eV.\cite{ray2015photoelectron} The optimized Ce-O bond lengths are 1.855~\AA\ for \ce{CeO^-} and 1.820~\AA\ for CeO, consistent with previous theoretical and experimental studies.\cite{kaledin1993laser,ray2015photoelectron}
\begin{table}[!ht]
\centering
\caption{Adiabatic and vertical detachment energies (ADE and VDE, in eV) for CeO$^{-}$/CeO computed at DFT-optimized ground-state geometries using different methods.} \label{tab:ADE_VDE}
\begin{tabular}{lcccc}
\toprule
Parameter & DFT$^{a}$ & CASSCF & NEVPT2 & Literature$^{b}$ \\
\midrule
ADE & 0.86 & 0.18 & 0.93 & 0.93 (exp.)/0.87 (calc.) \\
VDE & 0.87 & 0.25 & 0.84 & 0.89 \\
\bottomrule
\end{tabular}
\begin{tablenotes}
\footnotesize
\item[a] $^{a}$DFT values correspond to B3LYP-D3/ZORA/def2-TZVP calculations including electronic and zero-point vibrational contributions.
\item[b] $^{b}$Experimental and previously reported computed values from Ray et al. \textit{J. Chem. Phys.} \textbf{2015}, \textit{142}, 064305.
\end{tablenotes}
\end{table}
To obtain a more reliable description of the detachment energetics, multi-reference calculations have been performed for CeO/CeO$^-$. Table \ref{tab:ADE_VDE} reports the computed ADE and VDE values for \ce{CeO^-}. Inclusion of dynamic correlation through NEVPT2 produces a substantial shift in the detachment energetics relative to the CASSCF results, bringing the calculated values closer to experiment and supporting the use of NEVPT2 for subsequent spectroscopic analysis. 
\begin{figure}[!ht]
\centering
\includegraphics[width=10cm]{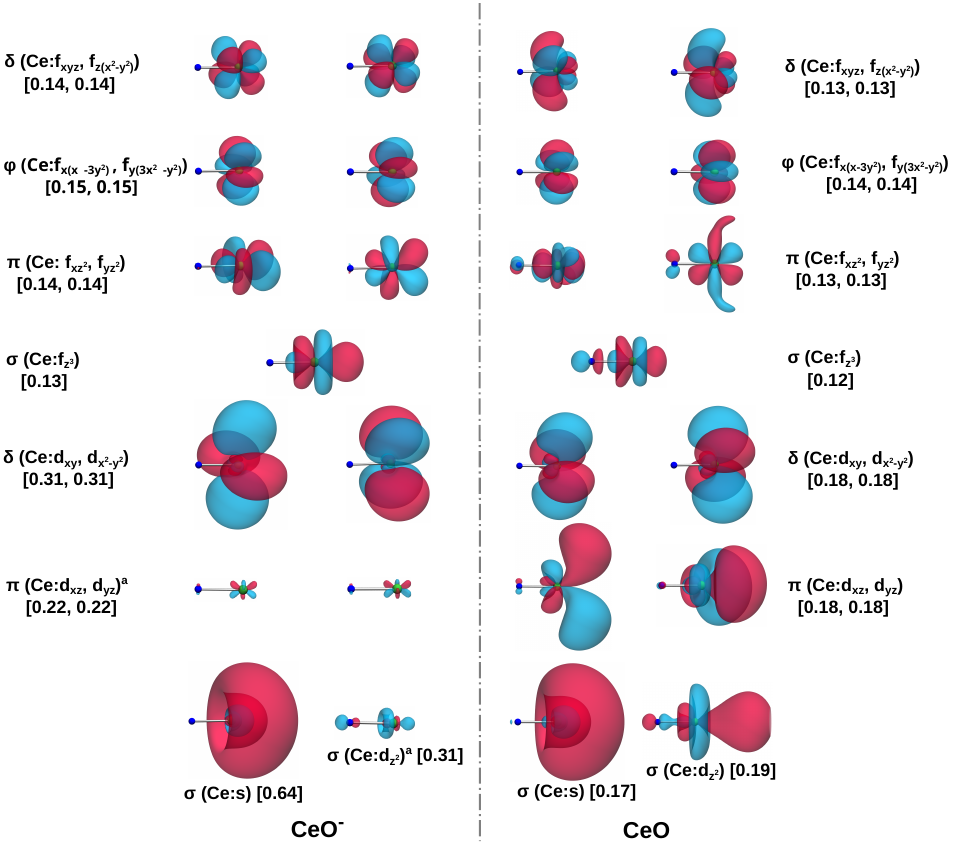}
 \caption{Active orbitals defining the CAS$(n,13)$ space for CeO$^-$ ($n = 3$) and CeO ($n = 2$), obtained from CASSCF+NEVPT2 calculations. The active space comprises Ce-centered $4f$, $5d$, and $6s$ orbitals that describe the near-degenerate valence manifold. Orbital isosurfaces are plotted at an isovalue of 0.03; numbers in brackets denote natural occupation numbers. Superscript $^{a}$ identifies the more diffuse Ce-centered $5d$ orbitals ($d_{z^2}$, $d_{xz}$, and $d_{yz}$), whose full spatial extent is not captured at 0.03.}

    \label{fig:cas_mos}
\end{figure}

\begin{figure}[!ht]
\centering
\includegraphics[width=14cm]{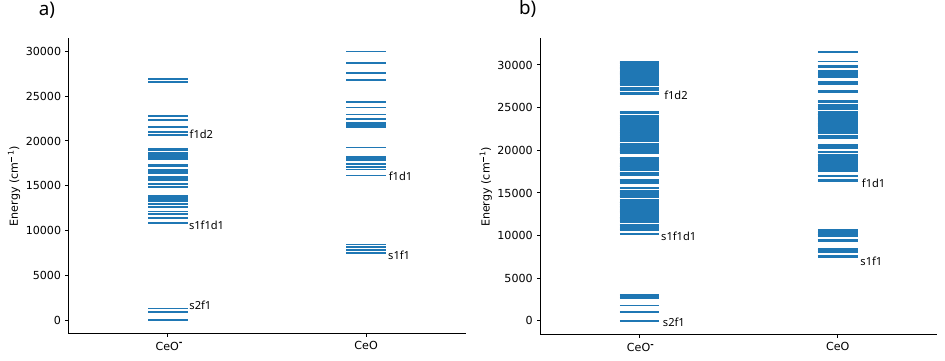}
\caption{Relative electronic term energies of CeO$^{-}$ (left) and CeO (right) computed using CASSCF+NEVPT2 method. Panel (a) shows spin–orbit–free results, while panel (b) includes spin–orbit coupling. All energies are referenced to the corresponding ground states.}
    \label{fig:levels}
\end{figure}

We performed spin-free and spin--orbit-coupled multireference calculations to characterize the low-lying electronic structure of CeO and \ce{CeO^-}. The active orbitals (Figure 1) reveal significant Ce \(6s\)/\(5d_{z^2}\) mixing in the anion and pronounced \(5d\)--\(4f\) hybridization in the neutral, while fractional occupations of the Ce \(4f\), \(5d\), and \(6s\) orbitals confirm the multiconfigurational character of both species. The resulting spin-free and spin--orbit-coupled excitation spectra are shown in Figure~\ref{fig:levels}. The anion exhibits a relatively sparse low-energy manifold dominated by the \(s^2f^1\) configuration, whereas the neutral displays a substantially higher density of states arising from competing \(s^1f^1\) and \(f^1d^1\) configurations. Spin--orbit coupling further increases the state density through splitting and mixing, particularly in the neutral. These electronic manifolds provide a framework for interpreting the \ce{CeO^-} photoelectron spectrum: near-threshold detachment populates low-lying \(s^1f^1\) states, while higher binding-energy features increasingly involve configurations with Ce \(5d\) character. At still higher energies, the dense \(f^1d^n\) manifold and strong spin--orbit interactions give rise to broad, unresolved features with \(\beta \approx 0\). Because the first vertical detachment energy (\(\sim 0.93\) eV) marks the onset of the continuum, these higher-lying states are best viewed as qualitative indicators of state density rather than strictly bound electronic states.

\section{$\Omega$-Level Structure in CeO}
\begin{figure}[!htbp]
\centering
\includegraphics[width=8cm]{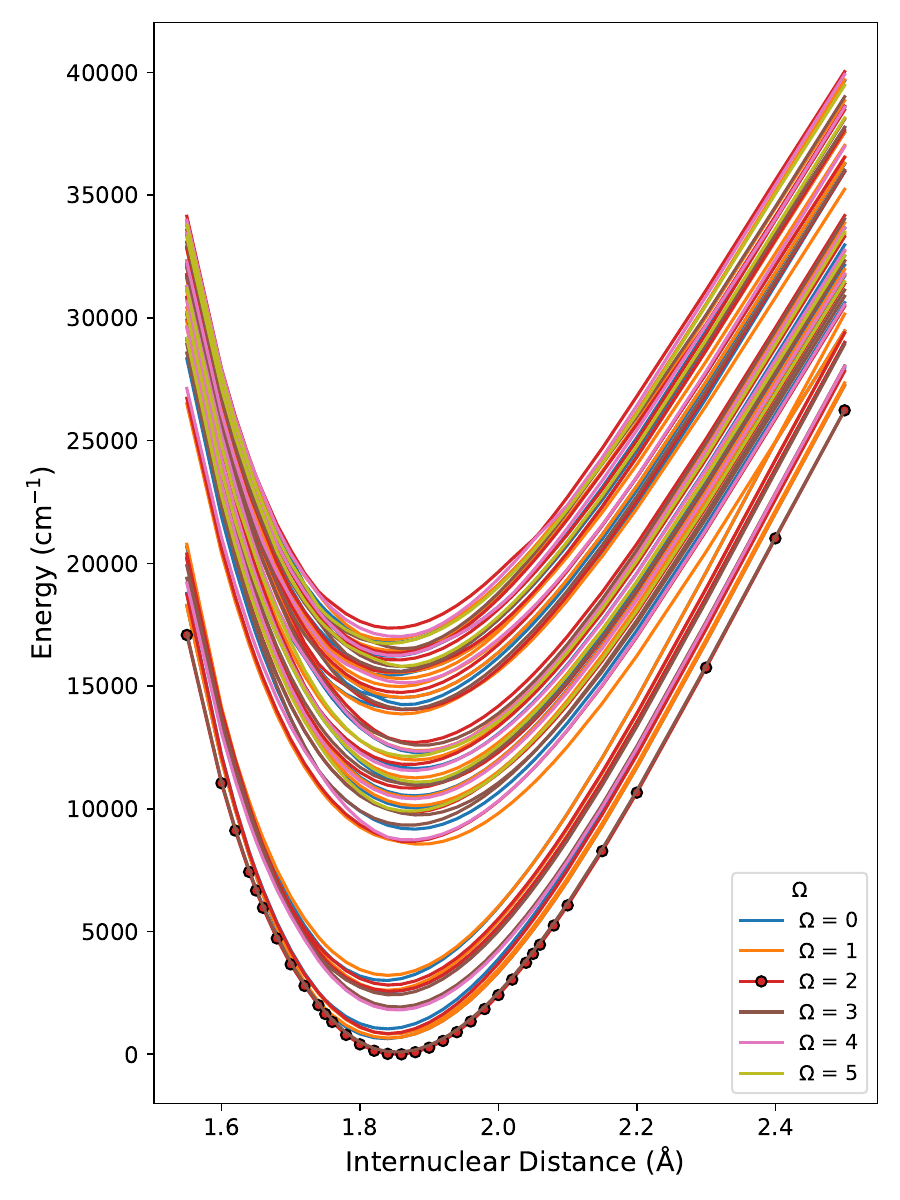}
\caption{ SOC potential energy curves of low-lying CeO states as a function of internuclear distance, labeled by \(\Omega\). The lowest manifold is dominated by \(s^1f^1\) configurations, while higher-lying states correlate primarily with \(f^1d^1\) configurations.}
    \label{fig:MJ_all_pec_soc}
\end{figure}

Table~\ref{tab:omega} summarizes the spin-orbit states of CeO, listing the total angular momentum projection (\(\Omega\)), excitation energies relative to the ground SO state (\(\Delta E\)), and dominant parent terms (\(\Lambda S\)). The state compositions reveal substantial multiconfigurational character and significant spin--orbit-induced singlet--triplet mixing, particularly within the low-lying \(\Phi\) and \(\Delta\) manifolds. 
\begin{table}[!ht]
\centering
\caption{Low-lying spin-orbit states of CeO at the CASSCF+NEVPT2 level with the CAS(2,13) active space. Energies are relative to the lowest state.}
\label{tab:omega}
\sisetup{table-format=1.3}
\begin{tabular}{l c S l p{6cm}}
\toprule
State & $\Omega$ & {$\Delta E$ (meV)} & $\Lambda S$ parent terms & Main composition \\
\midrule
0  & 2 & 0.000 & $^3\Phi$                 & 92\% $^3\Phi_2$ \\
2  & 3 & 0.092 & $^3\Phi$, $^1\Phi$       & 51\% $^3\Phi_3$ + 39\% $^1\Phi_3$ \\
4  & 0 & 0.529 & $^3\Sigma$, $^3\Pi$      & 59\% $^3\Sigma_{0}$ + 41\% $^3\Pi_{0}$ \\
5  & 1 & 0.576 & $^3\Sigma$, $^3\Pi$, $^3\Delta$ & 34\% $^3\Sigma_1$ + 32\% $^3\Pi_1$ + 30\% $^3\Delta_1$ \\
7  & 1 & 0.756 & $^3\Delta$, $^3\Sigma$, $^1\Pi$ & 48\% $^3\Delta_1$ + 30\% $^3\Sigma_1$ + 18\% $^1\Pi_1$ \\
9  & 2 & 0.761 & $^3\Delta$, $^1\Delta$   & 43\% $^3\Delta_2$ + 30\% $^1\Delta_2$ \\
11  & 0 & 0.936 & $^3\Pi$, $^1\Sigma$      & 64\% $^3\Pi_{0}$ + 35\% $^1\Sigma_{0}$ \\
12  & 4 & 1.813 & $^3\Phi$                 & 98\% $^3\Phi_4$ \\
14  & 3 & 1.932 & $^1\Phi$, $^3\Phi$       & 52\% $^1\Phi_3$ + 46\% $^3\Phi_3$ \\
16 & 3 & 2.371 & $^3\Delta$                & 88\% $^3\Delta_3$ \\
18 & 0 & 2.438 & $^3\Pi$, $^3\Sigma$  & 56\% $^3\Pi_0$ + 41\% $^3\Sigma_0$  \\
19 & 2 & 2.508 & $^3\Delta$, $^1\Delta$   & 51\% $^3\Delta_{2}$ + 25\% $^1\Delta_{2}$ \\
21 & 1 & 2.512 & $^3\Pi$, $^3\Sigma$   & 62\% $^3\Pi_1$ + 28\% $^3\Sigma_1$ \\
23 & 2 & 2.747 & $^3\Pi$, $^1\Delta$       & 56\% $^3\Pi_2$ + 42\% $^1\Delta_2$ \\
25 & 0 & 2.891 & $^1\Sigma$, $^3\Pi$      & 63\% $^1\Sigma_0$ + 36\% $^3\Pi_0$  \\
26 & 1 & 3.127 & $^1\Pi$                   & 78\% $^1\Pi_1$ \\
\bottomrule
\end{tabular}
\end{table}
Figure~\ref{fig:MJ_all_pec_soc}(a) illustrates the transformation of the low-lying CeO electronic structure upon inclusion of spin--orbit coupling. While the spin-free spectrum consists of well-defined \(^{3}\Phi\), \(^{1}\Phi\), \(^{3}\Delta\), \(^{1}\Delta\), \(^{3}\Pi\), \(^{1}\Pi\), \(^{3}\Sigma^{+}\), and \(^{1}\Sigma^{+}\) terms, spin--orbit coupling generates a dense manifold of closely spaced \(\Omega\) states. The calculated compositions indicate significant singlet--triplet and interterm mixing, particularly among the low-lying \(\Phi\), \(\Delta\), and \(\Pi\) states. These strongly mixed SOC states govern the observed spectroscopic properties and photodetachment pathways of CeO. The spin-free CASSCF+NEVPT2 spectroscopic constants of the low-lying states of \ce{CeO^-} and CeO are reported in Table~\ref{tab:spectroscopic_parameters}, with energies given relative to the \ce{CeO^-} ground state.

\begin{table}[!htbp]
\centering
\caption{Spin-free CASSCF+NEVPT2 spectroscopic constants for low-lying \ce{CeO^-} states below the neutral manifold and for neutral CeO. State labels are written as ${m}{}^{n}\!X$, where \(n\) is the spin multiplicity and \(m\) indicates the energetic ordering of states with the same X term symbol. $T_e$, $R_e$, and $\omega_e$ represent the electronic excitation energy (cm$^{-1}$), equilibrium bond distance (\AA), and harmonic vibrational constant (cm$^{-1}$), respectively. Energies are referenced to the \ce{CeO^-} ground state.}
\begin{tabular}{lccc|lccc}
\hline
State & $T_e$  &  $R_e$  & $\omega_e$  & State & $T_e$  &  $R_e$  & $\omega_e$  \\
\hline
 \multicolumn{4}{c|}{Doublets} & \multicolumn{4}{c}{Doublets} \\
 \hline
 ${1}^2\Phi$ &0 	& 1.8941 &	736 & ${1}^2\Phi$ &0 	& 1.8941 &	736 \\
${1}^2\Delta$ &	1034 &	1.8843	& 787 & ${1}^2\Delta$ &	1034 &	1.8843	& 787 \\
${1}^2\Sigma^+$ 	& 1068	& 1.8704 &	725 & ${1}^2\Sigma^+$ 	& 1068	& 1.8704 &	725 \\
${1}^2\Pi$ &	1420 &	1.8787 &	775& ${1}^2\Pi$ &	1420 &	1.8787 &	775\\
\hline
\multicolumn{4}{c|}{Triplets} & \multicolumn{4}{c}{Singlets} \\
\hline
${1}^3\Phi$  & 7535 & 1.8541 & 800  & ${1}^1\Phi$   & 7774 & 1.8541 & 784 \\
${1}^3\Sigma^+$  & 7992 & 1.8333 & 782 & ${1}^1\Delta$ & 8619 & 1.8456 & 741 \\
${1}^3\Delta$ & 8210 & 1.8465 & 740 & ${1}^1\Sigma^+$ & 8857 & 1.8319 & 811 \\
${1}^3\Pi$     & 8547 & 1.8417 & 761 & ${1}^1\Pi$   & 9288 & 1.8403 & 761 \\
${2}^3\Delta$  & 16101 & 1.8641 & 776 & ${2}^1\Sigma^+$ & 16547 & 1.8765 & 723 \\
${1}^3$H     & 16435 & 1.8864 & 726 & ${2}^1\Pi$    & 17210 & 1.8813 & 744 \\
${2}^3\Phi$   & 16873 & 1.8775 & 730 & ${2}^1\Delta$ & 17983 & 1.8689 & 715 \\
${2}^3\Pi$     & 17274 & 1.8737 & 757 & ${1}^1\Gamma$ & 18701 & 1.8911 & 718 \\
${1}^3\Gamma$  & 17684 & 1.8724 & 758 & ${2}^1\Phi$   & 19028 & 1.8817 & 742 \\
${2}^3\Sigma^+$ & 17802 & 1.8859 & 705 & ${1}^1\Sigma^-$ &	20318 &	1.8796	& 728 \\
${1}^3\Sigma^-$ & 17818 & 1.8735 & 739 & ${3}^1\Phi$	&21236 &	1.8620	& 873 \\
${3}^3\Pi$     & 18731 & 1.8930   & 758 & ${3}^1\Pi$ &	21535	&1.8732	& 662 \\
${3}^3\Sigma^+$ & 21612 & 1.8570 
& 725 & ${3}^1\Delta$	&21543	& 1.8558 &	757 \\
 ${3}^3\Delta$  & 21731& 1.8646 & 773 & ${1}^1$H &	22287&	1.8712&	773\\
${4}^3\Pi$   & 21781   & 1.8384 & 619 & ${2}^1\Gamma$ &	22674	& 1.8435	&724 \\

${2}^3\Gamma$ & 21882 & 1.8688 & 751 & ${3}^1\Sigma^+$ &	22921	& 1.8455 &	680\\
\hline
\end{tabular}
\label{tab:spectroscopic_parameters}
\end{table}

\clearpage
\subsection{Spin-free character of SOC states near experimental peaks}

\renewcommand{\arraystretch}{0.7}
\begin{longtable}{cccccc}
\caption{Dominant spin-free components of the spin-orbit-coupled (SOC) states closer in energy to the experimental photoelectron peaks (A-E). }
\label{tab:possible_assign_components}\\
\hline
State & $T_e$+I.E. (eV) & Spin-free components & \\
\hline
5 $\Omega$=1 &
1.99 &
$76\,\%\,{}^3\Delta_1\left|\sigma_d\delta_d\right\rangle
+13\,\%\,{}^3\Pi_1\left|\delta_d\pi_f\right\rangle
+8\,\%\,{}^1\Pi_1\left|\delta_d\pi_f\right\rangle$ \\

5 $\Omega$=2 & 2.00 & 
$57\,\%\,{}^3\Delta_2\left|\sigma_d\delta_d\right\rangle
+ 35\,\%\,{}^3\Phi_2\left|\pi_d\delta_d\right\rangle
+ 4\,\%\,{}^1\Delta_2\left|\delta_d\sigma_f\right\rangle$ \\

2 $\Omega$=4 & 2.01 &  
$96\,\%\,{}^3{\text{H}}_4\left|\delta_d\phi_f\right\rangle
+4\,\%\,{}^3\Gamma_4\left|\delta_d\delta_f\right\rangle$ \\

5 $\Omega$=0 & 2.07 &   
$57\,\%\,{}^1\Pi_0\left|\delta_d\delta_f\right\rangle
+22\,\%\,{}^3\Sigma_0\left|\delta_d\delta_f\right\rangle
+20\,\%\,{}^3\Pi_0\left|\delta_d\pi_f\right\rangle$ \\

4 $\Omega$=3 & 2.09 &  
$60\,\%\,{}^3\Delta_3\left|\sigma_d\delta_d\right\rangle
+32\,\%\,{}^3\Phi_3\left|\pi_d\delta_d\right\rangle
+7\,\%\,{}^3\Gamma_3\left|\delta_d\delta_f\right\rangle$ \\

5 $\Omega$=3 & 2.14 & 
$64\,\%\,{}^3\Gamma_3\left|\delta_d\delta_f\right\rangle
+ 20\,\%\,{}^3\Delta_3\left|\sigma_d\delta_d\right\rangle
+ 7\,\%\,{}^3\Phi_3\left|\pi_d\delta_d\right\rangle$\\
&& $+ 7\,\%\,{}^1\Phi_3\left|\delta_d\pi_f\right\rangle$ \\

6 $\Omega$=2 & 2.15 & 
$44\,\%\,{}^3\Phi_2\left|\pi_d\delta_d\right\rangle
+ 23\,\%\,{}^3\Delta_2\left|\sigma_d\delta_d\right\rangle
+ 24\,\%\,{}^3\Pi_2\left|\delta_d\pi_f\right\rangle$\\
&&  $+ 9\,\%\,{}^1\Delta_2\left|\delta_d\sigma_f\right\rangle$ \\

1 $\Omega$=5 & 2.16 &  
$93\,\%\,{}^3{\text{H}}_5\left|\delta_d\phi_f\right\rangle
+6\,\%\,{}^3\Gamma_5\left|\delta_d\delta_f\right\rangle$ \\

6 $\Omega$=0 & 2.17 &  
$68\,\%\,{}^3\Pi_0\left|\delta_d\pi_f\right\rangle
+24\,\%\,{}^3\Sigma_0\left|\delta_d\delta_d\right\rangle
+7\,\%\,{}^1\Sigma_0\left|(\delta_d)^2\right\rangle$ \\

6 $\Omega$=1 & 2.19 & 
$41\,\%\,{}^1\Pi_1\left|\delta_d\pi_f\right\rangle
+ 21\,\%\,{}^3\Pi_1\left|\delta_d\pi_f\right\rangle
+ 18\,\%\,{}^3\Delta_1\left|\sigma_d\delta_d\right\rangle$\\
&&  $+ 7\,\%\,{}^3\Pi_1\left|\delta_d\phi_f\right\rangle
+ 3\,\%\,{}^3\Sigma_1\left|\delta_d\delta_f\right\rangle
+ 3\,\%\,{}^3\Sigma_1\left|\delta_d\delta_d\right\rangle$ \\

3 $\Omega$=4 & 2.22 &  
$60\,\%\,{}^3\Phi_4\left|\pi_d\delta_d\right\rangle
+32\,\%\,{}^3\Gamma_4\left|\delta_d\delta_f\right\rangle
+6\,\%\,{}^1\Gamma_4\left|\delta_d\delta_d\right\rangle$ \\

7 $\Omega$=1 & 2.23 & 
$28\,\%\,{}^3\Sigma_1\left|\delta_d\delta_f\right\rangle
+ 28\,\%\,{}^3\Sigma_1\left|\delta_d\delta_d\right\rangle
+ 26\,\%\,{}^3\Pi_1\left|\delta_d\pi_f\right\rangle$\\
&&  $+ 13\,\%\,{}^1\Pi_1\left|\delta_d\pi_f\right\rangle
+ 2\,\%\,{}^3\Pi_1\left|\delta_d\phi_f\right\rangle$ \\

7 $\Omega$=0 & 2.24 & 
$56\,\%\,{}^3\Pi_0\left|\delta_d\pi_f\right\rangle
+ 30\,\%\,{}^1\Sigma_0\left|\delta_d\delta_f\right\rangle
+ 8\,\%\,{}^3\Pi_0\left|\delta_d\phi_f\right\rangle$\\
&&  $+ 5\,\%\,{}^3\Sigma_0\left|\delta_d\delta_f\right\rangle$ \\

7 $\Omega$=2 & 2.27 & 
$40\,\%\,{}^3\Pi_2\left|\delta_d\pi_f\right\rangle
+ 24\,\%\,{}^1\Delta_2\left|\delta_d\sigma_f\right\rangle
+ 20\,\%\,{}^3\Phi_2\left|\pi_d\delta_d\right\rangle$\\
&&  $+ 13\,\%\,{}^3\Delta_2\left|\sigma_d\delta_d\right\rangle$ \\

6 $\Omega$=3 & 2.29 &  
$60\,\%\,{}^3\Phi_3\left|\pi_d\delta_d\right\rangle
+24\,\%\,{}^3\Gamma_3\left|\delta_d\delta_f\right\rangle
+16\,\%\,{}^3\Delta_3\left|\sigma_d\delta_d\right\rangle$ \\

8 $\Omega$=0 & 2.30 &  
$47\,\%\,{}^3\Sigma_0\left|\delta_d\delta_d\right\rangle
+27\,\%\,{}^3\Pi_0\left|\delta_d\phi_f\right\rangle
+25\,\%\,{}^3\Pi_0\left|\delta_d\pi_f\right\rangle$ \\

1 $\Omega$=6 & 2.31 &  
$100\,\%\,{}^3{\text{H}}_{6}\left|\delta_d\phi_f\right\rangle$  \\

8 $\Omega$=1 & 2.33 & 
$50\,\%\,{}^3\Sigma_1\left|\delta_d\delta_f\right\rangle
+ 18\,\%\,{}^3\Sigma_1\left|\delta_d\delta_d\right\rangle
+ 11\,\%\,{}^3\Pi_1\left|\delta_d\phi_f\right\rangle$\\
&&  $+ 9\,\%\,{}^3\Pi_1\left|\delta_d\pi_f\right\rangle
+ 5\,\%\,{}^1\Pi_1\left|\delta_d\pi_f\right\rangle $ \\
\hline
5 $\Omega$=4 & 2.36 &  
$62\,\%\,{}^3\Gamma_4\left|\delta_d\delta_f\right\rangle
+28\,\%\,{}^3\Phi_4\left|\pi_d\delta_d\right\rangle
+4\,\%\,{}^1\Gamma_4\left|\delta_d\delta_d\right\rangle$ \\

9 $\Omega$=0 & 2.37 & 
$53\,\%\,{}^3\Sigma_0\left|\delta_d\delta_f\right\rangle
+ 23\,\%\,{}^3\Pi_0\left|\delta_d\pi_f\right\rangle
+ 22\,\%\,{}^3\Pi_0\left|\delta_d\phi_d\right\rangle$\\
&&  $+ 1\,\%\,{}^1\Sigma_0\left|\delta_d\delta_f\right\rangle$ \\

8 $\Omega$=2 & 2.39 & 
$53\,\%\,{}^1\Delta_2\left|\delta_d\sigma_f\right\rangle
+ 32\,\%\,{}^3\Pi_2\left|\delta_d\pi_f\right\rangle
+ 5\,\%\,{}^3\Delta_2\left|\sigma_d\delta_d\right\rangle$\\
&&  $+ 4\,\%\,{}^3\Pi_2\left|\delta_d\phi_f\right\rangle
+ 3\,\%\,{}^3\Delta_2\left|\pi_d\phi_f\right\rangle
+ 1\,\%\,{}^1\Delta_2\left|\pi_d\phi_f\right\rangle$ \\

9 $\Omega$=1 & 2.42 &  
$38\,\%\,{}^3\Sigma_1\left|\delta_d\delta_d\right\rangle
+28\,\%\,{}^3\Pi_1\left|\delta_d\pi_f\right\rangle
+25\,\%\,{}^1\Pi_1\left|\delta_d\pi_f\right\rangle$ \\

2 $\Omega$=5 & 2.44 &  
$92\,\%\,{}^3\Gamma_5\left|\delta_d\delta_f\right\rangle
+6\,\%\,{}^3{\text{H}}_5\left|\delta_d\phi_f\right\rangle$ \\

10 $\Omega$=0 & 2.45 & 
$68\,\%\,{}^3\Pi_0\left|\delta_d\phi_f\right\rangle
+ 19\,\%\,{}^3\Sigma_0\left|\delta_d\delta_d\right\rangle
+ 7\,\%\,{}^1\Sigma_0\left|(\delta_d)^2\right\rangle$\\
&&  $+ 4\,\%\,{}^3\Pi_0\left|\delta_d\pi_f\right\rangle$ \\

11 $\Omega$=0 & 2.46 &  
$68\,\%\,{}^3\Pi_0\left|\delta_d\phi_d\right\rangle
+20\,\%\,{}^3\Sigma_0\left|\delta_d\delta_f\right\rangle
+10\,\%\,{}^1\Sigma_0\left|\delta_d\delta_f\right\rangle$ \\

6 $\Omega$=4 & 2.46 &  
$85\,\%\,{}^1\Gamma_4\left|\delta_d\delta_d\right\rangle
+8\,\%\,{}^3\Phi_4\left|\pi_d\delta_d\right\rangle
+1\,\%\,{}^3\Gamma_4\left|\pi_d\phi_f\right\rangle$ \\

10 $\Omega$=1 & 2.46 & 
$65\,\%\,{}^3\Pi_1\left|\delta_d\phi_f\right\rangle
+ 10\,\%\,{}^3\Sigma_1\left|\delta_d\delta_f\right\rangle
+ 10\,\%\,{}^3\Sigma_1\left|\delta_d\delta_d\right\rangle$\\
&&  $+ 5\,\%\,{}^1\Pi_1\left|\delta_d\phi_f\right\rangle
+ 4\,\%\,{}^1\Pi_1\left|\delta_d\pi_f\right\rangle
+ 1\,\%\,{}^1\Pi_1\left|\pi_d\delta_d\right\rangle$ \\

7 $\Omega$=3 & 2.49 &  
$89\,\%\,{}^1\Phi_3\left|\delta_d\pi_f\right\rangle$ \\

9 $\Omega$=2 & 2.51 & 
$84\,\%\,{}^3\Pi_2\left|\delta_d\phi_f\right\rangle
+ 7\,\%\,{}^1\Delta_2\left|\delta_d\sigma_f\right\rangle
+ 3\,\%\,{}^3\Delta_2\left|\pi_d\phi_f\right\rangle$\\
&&  $+ 2\,\%\,{}^1\Delta_2\left|\pi_d\phi_f\right\rangle$ \\

11 $\Omega$=1 &
2.65 &
$30\,\%\,{}^3\Sigma_1\left|\pi_d\pi_d\right\rangle
+26\,\%\,{}^3\Pi_1\left|\pi_d\delta_f\right\rangle
+22\,\%\,{}^1\Pi_1\left|\pi_d\delta_d\right\rangle
+14\,\%\,{}^1\Pi_1\left|\delta_d\phi_f\right\rangle$ \\

12 $\Omega$=0 & 2.67 & 
$48\,\%\,{}^3\Pi_0\left|\pi_d\delta_f\right\rangle
+ 42\,\%\,{}^3\Sigma_0\left|\pi_d\pi_d\right\rangle$ \\

10 $\Omega$=2 & 2.67 &  
$47\,\%\,{}^1\Delta_2\left|\pi_d\phi_f\right\rangle
+27\,\%\,{}^3\Delta_2\left|\pi_d\phi_f\right\rangle
+12\,\%\,{}^3\Gamma_2\left|\pi_d\delta_f\right\rangle
+8\,\%\,{}^3\Pi_2\left|\delta_d\phi_f\right\rangle$ \\

8 $\Omega$=3 & 2.67 &   
$88\,\%\,{}^3\Gamma_3\left|\pi_d\phi_f\right\rangle
+4\,\%\,{}^3\Phi_3\left|\pi_d\delta_f\right\rangle$ \\

13 $\Omega$=0 & 2.70 &  
$78\,\%\,{}^1\Sigma_0\left|\delta_d\delta_d\right\rangle
+9\,\%\,{}^3\Sigma_0\left|\delta_d\delta_d\right\rangle
+5\,\%\,{}^3\Sigma_0\left|\pi_d\pi_d\right\rangle$ \\

14 $\Omega$=0 & 2.73 &
$76\,\%\,{}^3\Pi_0\left|\pi_d\delta_f\right\rangle
+ 17\,\%\,{}^1\Sigma_0\left|\pi_d\pi_f\right\rangle$ 
\\

12 $\Omega$=1 & 2.73 & 
$64\,\%\,{}^3\Delta_1\left|\pi_d\phi_f\right\rangle
13\,\%\,{}^3\Pi_1\left|\pi_d\delta_f\right\rangle
+12\,\%\,{}^1\Pi_1\left|\pi_d\delta_d\right\rangle$ \\

11 $\Omega$=2 & 2.76 &  
$80\,\%\,{}^3\Phi_2\left|\pi_d\delta_f\right\rangle
+ 8\,\%\,{}^1\Delta_2\left|\pi_d\pi_d\right\rangle$
    \\
    13 $\Omega$=1 & 2.79 &  
$41\,\%\,{}^1\Pi_1\left|\delta_d\phi_f\right\rangle
+ 10\,\%\,{}^3\Pi_1\left|\pi_d\delta_f\right\rangle$
    \\

7 $\Omega$=4 & 2.81 &  
$70\,\%\,{}^3\Gamma_4\left|\pi_d\phi_f\right\rangle
+18\,\%\,{}^1\Gamma_4\left|\pi_d\phi_f\right\rangle
+8\,\%\,{}^3\Phi_4\left|\pi_d\delta_f\right\rangle
+2\,\%\,{}^1\Gamma_4\left|\delta_d\delta_d\right\rangle$ \\

14 $\Omega$=1 & 2.83 & 
$ 24\,\%\,{}^3\Sigma_1\left|\pi_d\pi_d\right\rangle
+19\,\%\,{}^3\Pi_1\left|\pi_d\delta_f\right\rangle
+ 13\,\%\,{}^1\Pi_1\left|\delta_d\phi_f\right\rangle$\\
&&  $+16\,\%\,{}^3\Delta_1\left|\pi_d\phi_f\right\rangle
+ 11\,\%\,{}^1\Pi_1\left|\pi_d\delta_d\right\rangle
+ 7\,\%\,{}^3\Pi_1\left|\delta_d\phi_f\right\rangle$ \\

15 $\Omega$=0 & 2.84 &  
$20\,\%\,{}^3\Sigma_0\left|\pi_d\pi_d\right\rangle
+40\,\%\,{}^3\Pi_0\left|\pi_d\delta_f\right\rangle
+28\,\%\,{}^3\Pi_0\left|\pi_d\sigma_f\right\rangle
+12\,\%\,{}^1\Sigma_0\left|(\pi_d)^2\right\rangle$ \\

15 $\Omega$=1 & 2.86 & 
$25\,\%\,{}^3\Sigma_1\left|\pi_d\pi_d\right\rangle
+20\,\%\,{}^1\Pi_1\left|\pi_d\delta_d\right\rangle
+14\,\%\,{}^3\Pi_1\left|\pi_d\sigma_f\right\rangle$ \\
12 $\Omega$=2 & 2.86 &
$64\,\%\,{}^3\Pi_2\left|\pi_d\delta_f\right\rangle
+14\,\%\,{}^3\Delta_2\left|\pi_d\phi_f\right\rangle
+6\,\%\,{}^1\Delta_2\left|\pi_d\pi_d\right\rangle$\\
&&$+5\,\%\,{}^3\Delta_2\left|\pi_d\pi_f\right\rangle
+4\,\%\,{}^3\Pi_2\left|\pi_d\sigma_f\right\rangle$   \\

16 $\Omega$=0 & 2.86 &  
$60\,\%\,{}^3\Pi_0\left|\pi_d\sigma_f\right\rangle
+ 17\,\%\,{}^3\Sigma_0\left|\pi_d\pi_f\right\rangle
+ 10\,\%\,{}^1\Sigma_0\left|\pi_d\pi_f\right\rangle$    \\
9 $\Omega$=3 & 2.86 &  
$42\,\%\,{}^3\Phi_3\left|\pi_d\delta_f\right\rangle
+ 32\,\%\,{}^3\Delta_3\left|\pi_d\phi_f\right\rangle    
+ 12\,\%\,{}^1\Phi_3\left|\pi_d\delta_d\right\rangle$   \\
10 $\Omega$=3 & 2.89 &  
$64\,\%\,{}^3\Delta_3\left|\pi_d\phi_f\right\rangle
+ 22\,\%\,{}^3\Phi_3\left|\pi_d\delta_f\right\rangle$    \\

3 $\Omega$=5 & 2.89 &  
$92\,\%\,{}^1\text{H}_5\left|\delta_d\phi_f\right\rangle
+ 5\,\%\,{}^3\Gamma_5\left|\pi_d\phi_f\right\rangle$    \\

13 $\Omega$=2 & 2.92 & 
$27\,\%\,{}^1\Delta_2\left|\pi_d\delta_f\right\rangle
+24\,\%\,{}^3\Pi_2\left|\pi_d\sigma_f\right\rangle
+16\,\%\,{}^3\Pi_2\left|\pi_d\delta_f\right\rangle
+10\,\%\,{}^3\Delta_2\left|\pi_d\phi_f\right\rangle$ \\
17 $\Omega$=0 & 2.94 &
$52\,\%\,{}^3\Pi_0\left|\pi_d\sigma_f\right\rangle
+ 26\,\%\,{}^3\Sigma_0\left|\pi_d\pi_f\right\rangle
+ 2\,\%\,{}^1\Sigma_0\left|\delta_d\delta_d\right\rangle 
+ 2\,\%\,{}^1\Sigma_0\left|\pi_d\pi_d\right\rangle$ 
\\
8 $\Omega$=4 & 2.94 &  
$72\,\%\,{}^1\Gamma_4\left|\pi_d\phi_f\right\rangle
+ 22\,\%\,{}^3\Gamma_4\left|\pi_d\phi_f\right\rangle$
    \\
14 $\Omega$=2 & 2.96 &  
$35\,\%\,{}^3\Delta_2\left|\pi_d\phi_f\right\rangle   
+ 24\,\%\,{}^3\Pi_2\left|\pi_d\sigma_f\right\rangle
+ 18\,\%\,{}^1\Delta_2\left|\pi_d\phi_f\right\rangle $   \\

16 $\Omega$=1 & 2.97 & 
$ 33\,\%\,{}^3\Pi_1\left|\pi_d\sigma_f\right\rangle
+ 25\,\%\,{}^1\Pi_1\left|\pi_d\delta_d\right\rangle
+ 14\,\%\,{}^3\Sigma_1\left|\pi_d\pi_d\right\rangle$\\
&&  $+ 10\,\%\,{}^3\Sigma_1\left|\pi_d\pi_f\right\rangle
+8\,\%\,{}^3\Delta_1\left|\pi_d\pi_f\right\rangle
+ 4\,\%\,{}^1\Pi_1\left|\delta_d\phi_f\right\rangle
+ 3\,\%\,{}^1\Pi_1\left|\sigma_d\pi_d\right\rangle$ \\

11 $\Omega$=3 & 2.98 &  
$70\,\%\,{}^1\Phi_3\left|\delta_d\delta_d\right\rangle   
+ 21\,\%\,{}^3\Phi_3\left|\pi_d\delta_f\right\rangle
+ 8\,\%\,{}^3\Gamma_3\left|\pi_d\phi_f\right\rangle $   \\

4 $\Omega$=5 & 3.01 &  
$95\,\%\,{}^3\Gamma_5\left|\pi_d\phi_f\right\rangle   
+ 5\,\%\,{}^1{\text{H}}_5\left|\delta_d\phi_f\right\rangle $   \\

18 $\Omega$=0 & 3.01 &
$62\,\%\,{}^1\Sigma_0\left|\pi_d\pi_f\right\rangle
+ 24\,\%\,{}^3\Pi_0\left|\pi_d\sigma_f\right\rangle
+ 12\,\%\,{}^3\Pi_0\left|\pi_d\delta_f\right\rangle 
+ 1\,\%\,{}^3\Sigma_0\left|\pi_d\pi_f\right\rangle$ \\

17 $\Omega$=1 & 3.02 & 
$44\,\%\,{}^3\Delta_1\left|\pi_d\pi_f\right\rangle
15\,\%\,{}^1\Pi_1\left|\sigma_d\pi_d\right\rangle
+15\,\%\,{}^3\Pi_1\left|\pi_d\delta_f\right\rangle$ \\
9 $\Omega$=4 & 3.04 &
$84\,\%\,{}^3\Phi_4\left|\pi_d\delta_f\right\rangle
+ 8\,\%\,{}^1\Gamma_4\left|\pi_d\phi_f\right\rangle$ 
\\
15 $\Omega$=2 & 3.08 &  
$57\,\%\,{}^1\Delta_2\left|\pi_d\pi_d\right\rangle
+ 22\,\%\,{}^3\Delta_2\left|\pi_d\pi_f\right\rangle
+ 10\,\%\,{}^3\Phi_2\left|\pi_d\delta_f\right\rangle
+ 7\,\%\,{}^3\Pi_2\left|\pi_d\sigma_f\right\rangle$ \\

    \bottomrule

\end{longtable}

\begin{table}[ht]
\centering
\caption{Computed bright spin--orbit-coupled (SOC) resonance states of CeO$^{-}$ in the 2.00--2.33~eV energy range and their dominant configuration-state function (CSF) compositions. The corresponding parent spin-free term characters are provided in the table footnote.}
\label{tab:possible_resonance_states}
\begin{tabular}{c c}
\hline
 $E$ (eV) & Composition$^a$  \\
\hline
2.05 &
22\% $\sigma_s\delta_d\delta_f$ +
16\% $\sigma_s\delta_d\pi_f$ +
16\% $\sigma_s\delta_d\phi_f$ +
13\% $\sigma_d\delta_d\delta_f$ +
9\% $\sigma_d\delta_d\pi_f$ \\

2.11 &
32\% $\sigma_s\pi_d\phi_f$ $+$ +
22\% $\sigma_s\sigma_d\pi_f$ +
7\% $\sigma_s\sigma_d\delta_f$ +

\\

 2.12 &
19\% $\sigma_s\delta_d\delta_f$ +
15\% $\sigma_s\delta_d\phi_f$ +
10\% $\sigma_s\pi_d\phi_f$ +
9\% $\sigma_d\delta_d\delta_f$ +
8\% $\sigma_s\delta_d\pi_f$ \\

 2.13 &
64\% $\sigma_s\pi_d\phi_f$ 

\\

2.15 &
91\% $\sigma_s\pi_d\phi_f$ +
7\% $\sigma_s\sigma_d\pi_f$ 
\\

2.15 &
63\% $\sigma_s\pi_d\phi_f$ +
14\% $\sigma_s\sigma_d\pi_f$ 
\\
\hline
\end{tabular}\\
\footnotesize
$^a$Parent spin-free term characters are $^2\Sigma$ ($\sigma_s\delta_d\delta_f$, $\sigma_d\delta_d\delta_f$), $^2\Pi$ ($\sigma_s\delta_d\pi_f$, $\sigma_s\delta_d\phi_f$), $^2\Phi$ ($\sigma_d\delta_d\pi_f$), $^4\Delta$/$^4\Gamma$ ($\sigma_s\pi_d\phi_f$), $^4\Delta$/$^4\Gamma$ + $^2\Delta$/$^2\Gamma$ ($\sigma_s\pi_d\phi_f$), $^2\Pi$ ($\sigma_s\sigma_d\pi_f$), $^2\Delta$ ($\sigma_s\sigma_d\delta_f$).

\end{table}

\begin{table*}[!ht]
\centering
\caption{Calculated photodetachment transitions from the mixed bright spin--orbit-coupled (SOC) resonance of CeO$^-$ at 2.12 eV. Relative intensities ($I_{\rm rel}$) are normalized within each $\Omega$ manifold. The energies listed correspond to the neutral CeO final states.}
\label{tab:state2.12}

\scriptsize
\setlength{\tabcolsep}{2.8pt}
\renewcommand{\arraystretch}{1.05}

\begin{tabular}{cccccccccccc}
\toprule
\multicolumn{12}{c}{CeO$^-$ SOC resonance (2.12 eV)} \\
\midrule
\multicolumn{4}{c}{$\Omega=3/2$} &
\multicolumn{4}{c}{$\Omega=5/2$} &
\multicolumn{4}{c}{$\Omega=7/2$} \\
\cmidrule(lr){1-4}
\cmidrule(lr){5-8}
\cmidrule(lr){9-12}
State & $I_{\rm rel}$ & $E_{\rm CeO}$ (eV) & Orb. &
State & $I_{\rm rel}$ & $E_{\rm CeO}$ (eV) & Orb. &
State & $I_{\rm rel}$ & $E_{\rm CeO}$ (eV) & Orb. \\
\midrule

$5\Omega=1$ &0.53&1.99&$\sigma$:$\delta$ (64:36) &
$5\Omega=1$ &0.37&1.99&$\delta$ &
$5\Omega=1$ &0.37&1.99&$\delta$ \\

$4\Omega=3$ &0.29&2.09&$\delta$ &
$4\Omega=3$ &0.11&2.09&$\sigma$ &
$5\Omega=2$ &0.26&2.00&$\delta$ \\

$5\Omega=3$ &0.12&2.14&$\delta$ &
$5\Omega=3$ &1.00&2.14&$\sigma$ &
$2\Omega=4$ &0.84&2.01&$\sigma$ \\

$6\Omega=2$ &0.12&2.15&$\sigma$ &
$6\Omega=2$ &0.11&2.15&$\sigma$ &
$4\Omega=3$ &0.11&2.09&$\sigma$ \\

$6\Omega=1$ &0.53&2.19&$\sigma$ :$\delta$ (74:26) &
$6\Omega=1$ &0.11&2.19&$\delta$ &
$5\Omega=3$ &1.00&2.14&$\sigma$ \\

$7\Omega=1$ &0.65&2.23&$\sigma$ &
$7\Omega=2$ &0.16&2.27&$\sigma$ &
$6\Omega=2$ &0.11&2.15&$\delta$ \\

$7\Omega=2$ &0.18&2.27&$\sigma$ &
$6\Omega=3$ &0.37&2.29&$\sigma$ &
$6\Omega=1$ &0.11&2.19&$\delta$ \\

$6\Omega=3$ &0.06&2.29&$\delta$ &
&&&&
$3\Omega=4$ &0.47&2.22&$\sigma$ \\

$8\Omega=1$ &1.00&2.33&$\sigma$ &
&&&&
$7\Omega=2$ &0.05&2.27&$\delta$ \\

&&&&
&&&&
$6\Omega=3$ &0.37&2.29&$\sigma$ \\

\bottomrule
\end{tabular}
\end{table*}

\begin{figure}[!ht]
\centering
\includegraphics[width=10cm]{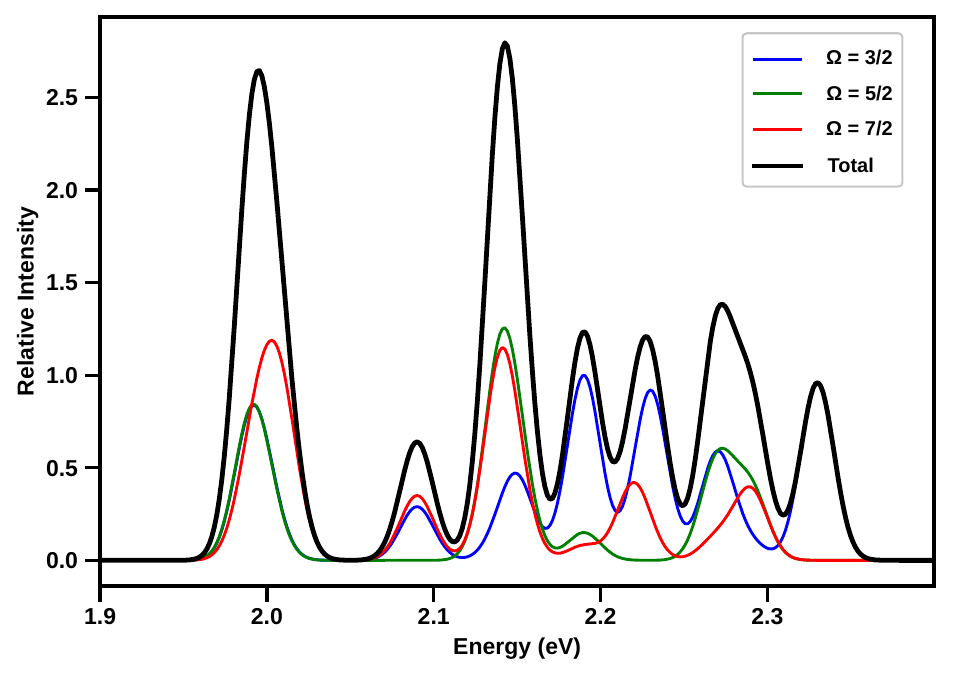}
\caption{Simulated photoelectron spectrum of CeO$^-$ obtained from the high-energy bright mixed spin--orbit coupled (SOC) resonance at 2.05~eV. The total spectrum is generated by applying Gaussian broadening (FWHM = 0.024~eV; $\sigma = 0.010$~eV) to the transitions to the neutral SOC states, weighted by the estimated relative photodetachment intensities derived from the wavefunction compositions. The individual contributions from the $\Omega=3/2$, $\Omega=5/2$, and $\Omega=7/2$ manifolds are also shown.}
   \label{fig:resonance_state
   2.05}
\end{figure}

\begin{table*}[ht]
\centering
\caption{Calculated photodetachment transitions from the mixed bright spin--orbit-coupled (SOC) resonance of CeO$^-$ at 2.05 eV. Relative intensities ($I_{\rm rel}$) are normalized within each $\Omega$ manifold. The energies listed correspond to the neutral CeO final states.}
\label{tab:state2.05}

\scriptsize
\setlength{\tabcolsep}{2.8pt}
\renewcommand{\arraystretch}{1.05}

\begin{tabular}{cccccccccccc}
\toprule
\multicolumn{12}{c}{CeO$^-$ SOC resonance (2.05 eV)} \\
\midrule
\multicolumn{4}{c}{ $\Omega=3/2$} &
\multicolumn{4}{c}{ $\Omega=5/2$} &
\multicolumn{4}{c}{ $\Omega=7/2$} \\
\cmidrule(lr){1-4}
\cmidrule(lr){5-8}
\cmidrule(lr){9-12}
State & $I_{\rm rel}$ & $E_{\rm CeO}$ (eV) & Orb. &
State & $I_{\rm rel}$ & $E_{\rm CeO}$ (eV) & Orb. &
State & $I_{\rm rel}$ & $E_{\rm CeO}$ (eV) & Orb. \\
\midrule

$5\Omega=1$ & 0.67 & 1.99 & $\sigma$:$\delta$ (70:30) &
$5\Omega=1$ & 0.69 & 1.99 & $\delta$:$\pi$ (54:46) &
$5\Omega=1$ & 0.38 & 1.99 & $\delta$ \\

$5\Omega=2$ & 0.25 & 2.00 & $\pi$ &
$5\Omega=2$ & 0.23 & 2.00 & $\pi$ &
$5\Omega=2$ & 0.54 & 2.00 & $\delta$:$\pi$ (54:46) \\

$4\Omega=3$ & 0.29 & 2.09 & $\pi$ &
$5\Omega=3$ & 1.00 & 2.14 & $\sigma$:$\pi$ (86:14) &
$2\Omega=4$ & 0.65 & 2.01 & $\sigma$ \\

$6\Omega=2$ & 0.42 & 2.15 & $\sigma$:$\pi$ (74:26) &
$6\Omega=2$ & 0.38 & 2.15 & $\sigma$:$\pi$ (74:26) &
$4\Omega=3$ & 0.35 & 2.09 & $\pi$:$\sigma$ (76:24) \\

$5\Omega=3$ & 0.08 & 2.14 & $\pi$ &
$6\Omega=1$ & 0.15 & 2.19 & $\delta$:$\pi$  (54:46) &
$5\Omega=3$ & 1.00 & 2.14 & $\sigma$:$\pi$ (86:14) \\

$6\Omega=1$ & 1.00 & 2.19 & $\sigma$:$\delta$ (70:30) &
$7\Omega=2$ & 0.54 & 2.27 & $\sigma$:$\pi$ (74:26) &
$6\Omega=2$ & 0.23 & 2.15 & $\delta$:$\pi$ (54:46) \\

$7\Omega=1$ & 0.92 & 2.23 & $\sigma$ &
$6\Omega=3$ & 0.38 & 2.29 & $\sigma$:$\pi$ (76:24) &
$6\Omega=1$ & 0.08 & 2.19 & $\delta$ \\

$7\Omega=2$ & 0.58 & 2.27 & $\sigma$:$\pi$ (74:26) &
&&&&
$3\Omega=4$ & 0.42 & 2.22 & $\sigma$ \\

$6\Omega=3$ & 0.08 & 2.29 & $\pi$ &
&&&&
$7\Omega=2$ & 0.12 & 2.27 & $\delta$:$\pi$ (54:46)\\

$8\Omega=1$ & 0.96 & 2.33 & $\sigma$ &
&&&&
$6\Omega=3$ & 0.38 & 2.29 & $\sigma$:$\pi$ (76:24) \\

\bottomrule
\end{tabular}
\end{table*}
\clearpage
\bibliography{Reference}